\documentclass[]{RAA}
\usepackage{graphicx,times}
\usepackage{epstopdf}
\usepackage{epsfig}

\begin{document}

\title{Comparison of Halo Detection from Noisy Weak Lensing Convergence Maps with Gaussian Smoothing and MRLens Treatment}

\volnopage{ {\bf 2010} Vol.\ {\bf XX} No. {\bf XX}, 000--000}
\setcounter{page}{1}

\author{Y.-X. Jiao, H.-Y. Shan and Z.-H. Fan
      \inst{}}
   \institute{Department of Astronomy, Peking University, Beijing 100871, China; {\it fanzuhui@pku.edu.cn}\\
\vs \no
   {\small Received [year] [month] [day]; accepted [year] [month] [day] }
}

\abstract{Taking into account the noise from intrinsic ellipticities
of source galaxies, we study the efficiency and completeness of halo
detections from weak lensing convergence maps. Particularly, with
numerical simulations, we compare the Gaussian filter with the so
called MRLens treatment based on the modification of the Maximum
Entropy Method. For a pure noise field without lensing signals, a
Gaussian smoothing results a residual noise field that is
approximately Gaussian in statistics if a large enough number of
galaxies are included in the smoothing window. On the other hand,
the noise field after the MRLens treatment is significantly
non-Gaussian, resulting complications in characterizing the noise
effects. Considering weak-lensing cluster detections, although the MRLens
treatment effectively deletes false peaks arising from noise,
it removes the real peaks heavily due to its inability to distinguish 
real signals with relatively low amplitudes from noise in its restoration process.
The higher the noise level is, the larger the removal effects are for the real peaks. 
For a survey with a source density $n_g\sim 30\hbox { arcmin}^{-2}$, 
the number of peaks found in an area of $3\times 3\hbox{ deg}^{2}$ after MRLens
filtering is only $\sim 50$ for the detection threshold $\kappa=0.02$, while the
number of halos with $M>5\times 10^{13}\hbox{ M}_{\odot}$ and with redshift $z\le 2$ 
in the same area is expected to be $\sim 530$. 
For the Gaussian smoothing treatment, the number of detections is $\sim 260$, much larger
than that of the MRLens. The Gaussianity of the noise statistics in the Gaussian smoothing case
adds further advantages for this method to circumvent the problem of the relatively low
efficiency in weak-lensing cluster detections. Therefore, in studies aiming to construct
large cluster samples from weak-lensing surveys, the Gaussian smoothing method performs
significantly better than the MRLens. 
}

\authorrunning{Y.-X. Jiao, H.-Y. Shan \& Z.-H. Fan}
\titlerunning{Comparison of Halo Detection}
\maketitle

\section{Introduction}
\label{sect:intro}

The weak gravitational lensing effect provides a unique tool in
measuring the matter distribution in the universe (e.g., Bartelmann
\& Schneider 2001; Hoekstra et al. 2006; Massey et al. 2007). Its
additional dependence on the distances to the source, to the lens
and between the source and lens makes it an excellent probe in
cosmological studies of dark energy (e.g., Albrecht et al. 2006;
Benjamin et al. 2007; Kilbinger et al. 2009; Li et al. 2009). On the
other hand, however, different observational and physical effects
can affect the weak lensing analyses significantly. Being extracted
from shape distortion of background galaxies, the weak lensing
effect on individual source galaxies is severely contaminated by
their intrinsic ellipticities. Therefore statistical analyses on a
large number of galaxies are necessary in weak lensing studies. Even
so, intrinsic shape alignments of galaxies, including
intrinsic-intrinsic and shear-intrinsic correlations, can be an
important source of error in cosmic shear correlation analyses. For
cluster detections from weak lensing convergence maps reconstructed
from shear measurements (e.g., Kaiser \& Squires 1993; Bartelmann
1995; Kaiser 1995; Schneider \& Seitz 1995; Squires \& Kaiser 1996;
Bridle et al.1998; Marshall et al. 2002), even randomly orientated
intrinsic ellipticities can result false peaks by their chance
alignments, which can reduce the efficiency of cluster detections
significantly (e.g., Schneider 1996; van Waerbeke 2000; White et al.
2002; Hamana et al. 2004; Fan 2007). Thus further treatments for a
convergence map are normally required to suppress the noise effects.

The noise from intrinsic ellipticities of source galaxies is
essentially shot noise, and thus by averaging over a relatively
large number of source galaxies in weak lensing analyses, the
residual noise can be effectively reduced. This leads to the normal
smoothing treatment. It is clear that the residual noise depends on
the form of the window function and the smoothing scale. For a
Gaussian smoothing with a window function of the form
$W(\theta)\propto \exp (-\theta^2/\theta_G^2)$, the residual noise
can be estimated by $\sigma^2_{0}\approx {(\sigma^2_{\epsilon}/
2)}{[1/( 2\pi \theta_G^2 n_g)]}$, where $\sigma_{\epsilon}$ is the
rms of the intrinsic ellipticity of individual source galaxies,
$\theta_G$ is the smoothing scale, and $n_g$ is the surface number
density of source galaxies. For $\sigma_{\epsilon}=0.3$,
$n_g=30\hbox{ arcmin}^{-2}$ and $\theta_G=1\hbox{ arcmin}$, we have $\sigma_{0}\approx 0.015$.

Recently, Starck et al. (2006) proposed the MRLens filtering
technique, which is based on the Bayesian analyses with a
multi-scale entropy prior applied. The False Detection Rate (FDR)
method is used to select significant/non-significant wavelet
coefficients (e.g., Starck et al. 2006; Pires et al. 2009). The
MRLens method suppresses noise adaptively according to the strength
of the noise itself. A more detailed description of the method is
given in \S4.

In this paper, with numerical simulations, we compare Gaussian
smoothing with MRLens treatment, paying particular attention to the
completeness and the efficiency of weak lensing halo detections from
convergence maps. The rest of the paper is organized as follows. In
\S2, we describe briefly the weak-lensing convergence reconstruction
and the Gaussian smoothing. In \S3, we present the important aspects
of the MRLens treatment. Results are shown in \S4. Section~5
contains summaries and discussions.

\section{Weak lensing convergence reconstruction}
\label{sect:rec}

In the weak lensing regime, the convergence $\kappa(\vec \theta)$ is
essentially related to the weighted projection of density
fluctuations $\delta$ along the unperturbed light path.
Specifically, we have
\begin{equation}
\kappa(\vec \theta)={3H_0^2\Omega_0\over 2}\int_0^{w_H} dw \bar{W}(w)f_K(w)
{\delta[f_K(w)\vec \theta,w]\over a(w)}
\end{equation}
where $H_0$ is the present Hubble constant, $\Omega_0$ is the
present matter density of the universe in unit of the critical
density, ${w}$ is the radial coordinate, $a(w)$ is the scale factor
of the universe, and, with $K$ being the spatial curvature of the universe,
\begin{eqnarray}
f_K(w)&& =|K|^{-1/2}\sin (|K|^{1/2}w)\quad \quad \ \ \   (K>0) \nonumber \\
&&=w \qquad \qquad \qquad \qquad \qquad \quad (K=0) \nonumber \\
&&=|K|^{-1/2}\sinh (|K|^{1/2}w) \qquad (K<0)\ .
\end{eqnarray}
The factor $\bar {W}(w)$ is the weighting function that is related to the
source galaxy distribution $G(w)$ by
\begin{equation}
\bar{W}(w)=\int_w^{w_H} dw' G(w'){f_K(w'-w)\over f_K(w')}\ .
\end{equation}

The lensing potential $\phi$ is related to $\kappa$ by
\begin{equation}
\kappa={\nabla^2\phi\over 2}\ ,
\end{equation}
and the shears $\gamma_1$ and $\gamma_2$ are
\begin{equation}
\gamma_1={\partial_{11}\phi - \partial_{22}\phi\over 2}\ , \quad \gamma_2=\partial_{12}\phi.
\end{equation}

Since both $\kappa$ and $\gamma_{i}$ are determined by the lensing
potential, they are mutually dependent of each other. In the Fourier
space, we have (Kaiser \& Squires 1993)
\begin{equation}
\kappa(\vec k)=c_{1}(k)\gamma_{1}(\vec k)+c_{2}(k)\gamma_{2}(\vec k),
\end{equation}
where $[c_1, c_2]=[\cos(2\phi),\sin(2\phi)]$ with $\vec k=k(\cos \phi, \sin \phi)$.

Observationally, the shear $\gamma$ can be extracted from the shape
measurement of source galaxy images. Under the condition
$\kappa<<1$, we have
\begin{equation}
\vec e^{\rm obs} \approx \vec \gamma+\vec e^{S}\ ,
\end{equation}
where $\vec e^{\rm obs}$ and $\vec e^{S}$ are the observed ellipticity, and
the intrinsic ellipticity of a source galaxy, respectively.
Reconstructed from $\vec e^{\rm obs}$, the convergence $\kappa_n(\vec k)$ then
contains noise from the intrinsic part, i.e.,
\begin{equation}
\kappa_n({\vec  k})=c_{\alpha}(k)e^{\rm obs}_{\alpha} ({\vec k})=
\kappa({\vec k})+c_{\alpha}(k) e^{S}_{\alpha}({\vec k}).
\end{equation}
With the transformation back to the real 2-D space and applying a
smoothing with the window function $W(\vec \theta)$, we can obtain the
smoothed quantities (e.g., van Waerbeke 2000)
\begin{equation}
\Sigma^{\rm
obs}(\vec \theta)=\Gamma(\vec \theta)+\frac{1}{n_g}\Sigma^{N_g}_{i=1}
W(\vec \theta-\vec \theta_i)e^{\rm S}(\vec \theta_i)
\end{equation}
and
\begin{equation}
K_N(\vec \theta)=\int d\vec k e^{-i\vec k \cdot \vec \theta}c_{\alpha}(k)\Sigma_{\alpha}^{\rm obs}(\vec k),
\end{equation}
where $\Sigma^{\rm obs}$, $\Gamma$, and $K_N$ are the
smoothed $e^{\rm obs}$, $\gamma$ and $\kappa_n$, respectively, and
$n_g$ and $N_g$ are the surface number density and the total number
of source galaxies in the field. The noise part of $K_N$ due to the
intrinsic ellipticities is then
\begin{equation}
N(\vec \theta)=\frac{1}{n_g}\Sigma^{N_g}_{i=1}\int d\vec k W(\vec k)
e^{-i\vec k\cdot (\vec \theta-\vec \theta_i)}c_{\alpha}(k)e^{S}_{\alpha}(\vec \theta_i),
\end{equation}
where $W(\vec k)$ is the Fourier transformation of the window function
with the form
\begin{equation}
W(\vec k)=\frac{1}{(2\pi)^2}\int d\vec \theta e^{i\vec k\cdot \vec \theta}W(\vec \theta).
\end{equation}

Without considering the intrinsic alignment of $e^{S}$, it is
expected from the central limit theorem that the smoothed noise
field $N(\vec \theta)$ is approximately Gaussian in statistics if the
effective number of galaxies included in the smoothing window is
larger than about $10$ (e.g., van Waerbeke 2000). In this case,
smoothing leads to correlations in $N(\vec \theta)$, and its two-point
correlation function is approximately
\begin{equation}
<N(\vec \theta)N(\vec \theta')>=\frac{\sigma_{\epsilon^2}}{2n_g}(2\pi)^2
\int d\vec k e^{i\vec k\cdot (\vec \theta'-\vec \theta)}|W(\vec k)|^2,
\end{equation}
where $\sigma_{\epsilon}$ is the intrinsic dispersion of $e^{\rm
obs}$.

The approximate Gaussianity of $N(\vec \theta)$ allows us to quantify the
noise effects straightforwardly. The noise effects on cluster mass
reconstruction and the noise peak statistics are analyzed in van
Waerbeke (2000). Even with weak alignments of intrinsic
ellipticities, $N(\vec \theta)$ can still be approximately described by a
Gaussian random field with a modified two-point correlation function
including the effects of intrinsic alignments. The enhancement of
the noise peak abundance due to the weakly intrinsic alignments are
analyzed in Fan (2007). In Fan et al. (2010), the effects of the
presence of real dark matter halos on the noise peak statistics
around them as well as the effects of the noise on the peak height
of real halos are investigated in detail. They further present a
model to calculate the total peak abundance in a large-scale
convergence map, including the peaks corresponding to real halos and
the noise peaks from the chance alignment of the intrinsic
ellipticities of source galaxies. Such a model makes it possible for
us to use directly the peaks from convergence maps as cosmological
probes without the need to differentiate real and false peaks.

Due to its simple operational procedure and the Gaussian statistics
of the residual noise field, the smoothing treatment has been widely
applied in weak lensing analyses. Different smoothing functions have
been used in different studies. In this paper, we consider the
Gaussian smoothing function $W_G$, which is one of the most commonly
adopted window functions. Specifically, we have
\begin{equation}
W_G=\frac{1}{\pi \theta_G^2} \exp
\left(-\frac{\theta^2}{\theta_G^2} \right)\ ,
\end{equation}
where $\theta_G$ is the smoothing scale.
Then from Eq.~(13), the rms of the noise $\sigma_0$ after smoothing
is given by
\begin{equation}
\sigma_0^2={\sigma_{\epsilon}^2\over 2}{1\over 2\pi \theta_G^2 n_g}\ .
\end{equation}
In our analyses, we choose $\sigma_{\epsilon}=0.3$, the typical
value for lensing source galaxies, and $\theta_G=1\hbox{ arcmin}$,
which is the optimal smoothing scale considering cluster-sized
halos. Then for a lensing survey with $n_g=30\hbox{ arcmin}^{-2}$,
$\sigma_0\approx 0.015$, which is about $20$ times lower than
$\sigma_{\epsilon}$.

\section{MRLens method}
\label{sect:MRLens}

Starck et al. (2006) introduce a new reconstruction and filtering
method, namely, Multi-scale Entropy Restoration (MRLens). It is
developed from the Maximum Entropy Method. The basic idea is to use
only `signals' selected by the so called False Discovery Rate (FDR)
(Benjamini \& Hochberg 1995) to reconstruct the convergence field
through a Multi-scale Entropy prior. In the following, we present
specific steps of MRLens.

\subsection{Wavelet decomposition}

For an original convergence map $\kappa_{obs}$ with $N=n\times n$ pixels, the
first step of MRLens is to decompose the image map into different
components representing fine structures of different scales.

To do this, we first initialize $j=0$ and set $C_0(k,l)=\kappa_{obs}(k,l)$,
i.e., $j=0$ corresponds to the unprocessed map with detailed
structures. Then we progressively go to higher $j$ to obtain
smoother maps through (Starck et al. 2001)

\begin{equation}
C_{j+1}(k,l)=\sum_{m}\sum_{n}h_{1D}(m)h_{1D}(n)C_{j}(k+2^{j}m,l+2^{j}n),
\end{equation}
where $h_{1D}(m)={[1/16,4/16,6/16,4/16,1/16]}$ for $m=-2,-1,0,1,2$,
respectively. Defining
\begin{equation}
w_{j+1}(k,l)=C_j(k,l)-C_{j+1}(k,l),
\end{equation}
we finally obtain
\begin{equation}
\kappa_{obs}(k,l) = C_{J}(k,l) + \sum_{j=1}^{J} w_{j}(k.l),
\end{equation}
where $J$ is a chosen number determined by specific considerations
on how smooth we want to go. Here we set $J=7$. In our following
analyses, each map is $3\times 3 \hbox{ deg}^2$ discretized into
$1024\times 1024$ pixels. Thus $2^J=128$ pixels corresponding to
$\sim 22 \hbox{ arcmin}$. Because we do not expect to see
significant structures resulting purely from noise on such a large
scale, $J=7$ is an appropriate choice.

It can be seen from Eq.~(18) that $C_{J}(k,l)$ is the most smoothed
version of the original map $\kappa_{obs}$, and the terms in the summation
contain ever smaller-scale information with smaller $j$.

\subsection{Multiscale Entropy}

With the multi-scale wavelet decomposition, one can then construct
an entropy with the obtained wavelet coefficients $w_{j}(k,l)$ at
each grid $(k,l)$ with $j=1,2,...,J$. It can generally be written as
\begin{equation}
H(\kappa) =\sum_{k,l}h[C_J(k,l)]+\sum_{j=1}^{J}\sum_{k,l}h[w_j(k,l)].
\end{equation}
For $h$, there are different definitions (e.g., Starck et al. 2006).

Here we follow Starck et al. (2001) to choose the entropy of
NOISE-MSE $h_n$ in our considerations. At each scale $j$, the noise
entropy at each grid $(k,l)$ is derived by weighting the entropy
with a probability that $w_j(k,l)$ is contributed by noise.
Specifically, we have
\begin{equation}
h_n[w_{j}(k,l)] =  \int_{0}^{\mid w_{j}(k,l)\mid } P_n[\mid w_{j}(k,l)
\mid -u]\frac{\partial h(x)}{\partial x}|_{x=u}du,
\end{equation}
where $P_n[w_{j}(k,l)]$ is the probability that the coefficient
$w_{j}(k,l)$ can be due to noise, and is given by
\begin{equation}
 P_n[w_{j}(k,l)] =\mathrm{Prob}[W > \mid w_{j}(k,l) \mid].
\end{equation}
Eq.~(20) essentially regards the information contained in $w_j(k,l)$
to be built up from the summation of $dh(u)$. For each newly added
$dh(u)$, depending on the difference $|w_j(k,l)|-u$, there is a
probability that it is due to noise.

For Gaussian noise with rms $\sigma_j$ at scale $j$, we have
\begin{eqnarray}
 P_n[w_j(k,l)] & =  & \frac{2}{\sqrt{2 \pi} \sigma_j}
 \int_{\mid w_j(k,l) \mid}^{+\infty} \exp(-W^2/2\sigma^2_j) dW \nonumber \\
 & = & \mbox{erfc}\bigg (\frac{\mid w_{j,k,l} \mid }{\sqrt{2}\sigma_j}\bigg )
\end{eqnarray}
and thus
\begin{equation}
h_n[w_{j}(k,l)] = \frac{1}{\sigma_j^2} \int_{0}^{\mid w_{j}(k,l) \mid} u
              \mbox{ erfc}\bigg (\frac{\mid w_{j}(k,l) \mid -u}{\sqrt{2}
              \sigma_j}\bigg )du.
\end{equation}

\subsection{Selecting significant wavelet coefficients using the False Discovery Rate (FDR)}

The Multiscale Entropy method applies regularizations on wavelet
coefficients to minimize noise contributions while keeping the
signal information. Thus for those coefficients which are clearly
signals, they should be kept unchanged. Then a new Multiscale
Entropy is defined as (e.g., Starck et al. 2006)
\begin{equation}
\tilde h_{n}[w_{j}(k,l)]=\bar M_{j}(k,l)h_n[w_{j}(k,l)],
\end{equation}
where
\begin{equation}
\bar M_{j}(k,l)=1-M_{j}(k,l),
\end{equation}
and $M$ is the multi-resolution support defined as (Starck et al.
1995)
\begin{equation}
M_{j}(k,l)=\{\begin{array}{ll} 1 & \textrm{if $w_{j}(k,l)$ is significant}\\ 0 & \textrm{if $w_{j}(k,l)$ is not significant}\end{array}.
\end{equation}
Therefore $\tilde h_n$ means that we only need to regularize those
wavelet coefficients which are $^\prime$not significant$^\prime$,
that is, they are likely due to noise.

For judging the significance of a wavelet coefficient, a commonly
used criterion is a `$k\sigma$' threshold. If a coefficient is above
the threshold, it is defined to be $^\prime$significant$^\prime$.
This is equivalent to set a threshold for the ratio of
$^\prime$significant$^\prime$ detections over the total number of
pixels being analyzed. Considering a Gaussian noise, a $2\sigma$
criterion corresponds to a probability of $0.05$ for a noise
coefficient being mis-classified as $^\prime$significant$^\prime$.
If we have totally $N$ pixels to consider, the number of false
discoveries is then on average $0.05N$. If the number of pixels
related to real signals in the analyses is comparable to $0.05N$,
the false discovery rate with respect to the number of real signals
can be much higher than $0.05$. Increasing $k$ can lower the number
of false detections at the expense, however, of the power of real
detections. To overcome such difficulties, an alternative
thresholding technique, the False Discovery Rate (FDR), has been
proposed (Benjamini \& Hochberg 1995; Miller et al. 2001; Hopkins et
al. 2002; Starck et al. 2006).

This method can effectively control, in an adaptive manner, the
fraction of false discoveries over the total number of discoveries,
rather than over the total number of pixels analyzed.

Let $P_1, \ldots, P_N$ denote the p-values ordered from low to high
for the N pixels, where p-value is defined as
\begin{equation}
p_{value}=\frac{1}{\sqrt{2\pi}\sigma_{j}}\int_{w_{j}(k,l)}^{\infty}\exp[-(w- \bar w_{j})^2/2\sigma^2_{j}]dw,
\end{equation}
with $\bar w_{j}$ the average of $w_{j}(k,l)$ for the scale $j$ over all the pixels.
Define
\begin{equation}
d_j =\max\left\{ k_j :\ P_{k_j} < \frac{k_j\alpha_j}{c_{N} N} \right\},
\end{equation}
then all the $w_{j}(k,l)$ with their values larger than $d_j$ are
classified as $^\prime$significant$^\prime$. Here $c_{N}=1$ if all
the pixels are statistically independent. The meaning of $\alpha_j$
is approximately the pre-defined false discovery rate at scale $j$
with respect to the total number of detections. The larger the
$\alpha_j$ value, the larger the fraction of $w_{j}(k,l)$ defined to
be $^\prime$significant$^\prime$. In our analyses, we adopt FDR to
find the values of $M$ in Eq.~(26). In the MRLens program,
$\alpha_0$ is an adjustable parameter, and $\alpha_j=\alpha_0\times
2^j$ (Starck et al. 2006).

\subsection{Multi-scale Entropy Filtering algorithm}

Given the discussions in the previous subsections, the Multi-scale
Entropy restoration method reduces to find the reconstructed
$\kappa_{f}$ that minimizes $I(\kappa_{f})$ defined as
\begin{equation}
I({\kappa_{f}})=\frac{\parallel{\kappa_{obs}-\kappa_{f}\parallel^2}}{2\sigma_n^2}+\beta\sum_{j=1}^{J}\sum_{k,l}\tilde h_n[({\cal
W}{\kappa_{f}})_{j}(k,l)],
\end{equation}
where $\sigma_{n}$ is the rms of noise in the original convergence
map $\kappa_{obs}$, $J$ is the number of wavelet scales, $\cal W$ is
the wavelet transform operator and $\tilde h_n[({\cal
W}{\kappa_{f}})_{j}(k,l)]$ is the multi-scale entropy defined only
for non-significant coefficients selected by the FDR method. The
$\beta$ parameter is calculated under the restriction that the
residual should have a standard deviation equal to the rms of noise.
The best $\kappa_{f}$ is then obtained by iterative calculations.
Full details of the minimization algorithm can be found in Starck et
al. (2001).

It can be seen that the two terms in the right of Eq.~(29) are
balancing each other. While the first term tends to keep the information in $\kappa_f$ the most, 
the second term has the effect to lower the
noise as much as possible.

\section{Results}
\label{sect:Sim}

In this section, we present the results of our analyses. For
weak-lensing effects from large-scale structures in the universe, we
use the publicly available ray-tracing weak-lensing maps provided
kindly by White \& Vale (2004). The specific set of lensing maps we
analyze are generated from large-scale N-body simulations with
cosmological parameters $\Omega_M=0.296$, $\Omega_{\Lambda}=0.704$,
$w=-1.0$, $h=0.7$, and $\sigma_8=0.93$. The box size is $300 \hbox{
Mpc}~h^{-1}$, the number of particles is $512^3$ with $m \approx
1.7\times10^{10}M_{\odot}~h^{-1}$ for each, and the softening length
is $\approx 20\hbox{ kpc}~h^{-1}$. There are totally $16$
convergence maps and each has a size of $3\times 3\hbox{ deg}^2$
pixelized into $1024\times 1024$ pixels. The redshift distribution 
of source galaxies follows $p(z)\propto z^2\exp[-(z/z_0)^{3/2}]$ with $z_0=2/3$.

For each map, we add in Gaussian noise due to the intrinsic ellipticities
of source galaxies with the variance given by (e.g., Hamana et al. 2004),
\begin{equation}
\sigma_{\rm pix}^2=\frac{\sigma_{\epsilon}^2}{2} \frac{1}{n_g \theta_{\rm pix}^2},
\end{equation}
where $\theta_{\rm pix}$ is the pixel size of the simulated
convergence-$\kappa$ map, and $\sigma_{\epsilon}$ is the rms of the
intrinsic ellipticites taken to be $\sigma_{\epsilon}=0.3$. The
surface number density $n_g$ depends on specific observations. Here
we consider $n_g=30\hbox{ arcmin}^{-2}$ which is typical for
ground-based observations, and $n_g=100 \hbox{ arcmin}^{-2}$ expected
from space observations, respectively. Figure~1 presents one set of
convergence maps without (left) and with (right) noise. It can be
seen very clearly that the noise from intrinsic ellipticities of
source galaxies dominates the map, and certain post-processing
procedures are necessary in order to extract weak-lensing signals
embedded under noise. Here we compare two such methods, namely, the
normal smoothing method with a Gaussian smoothing function, and the
MRLens treatment, paying particular attention to their effects on
weak-lensing peak statistics.

\begin{figure}[h!!!]
   \includegraphics[width=7.5cm, angle=0]{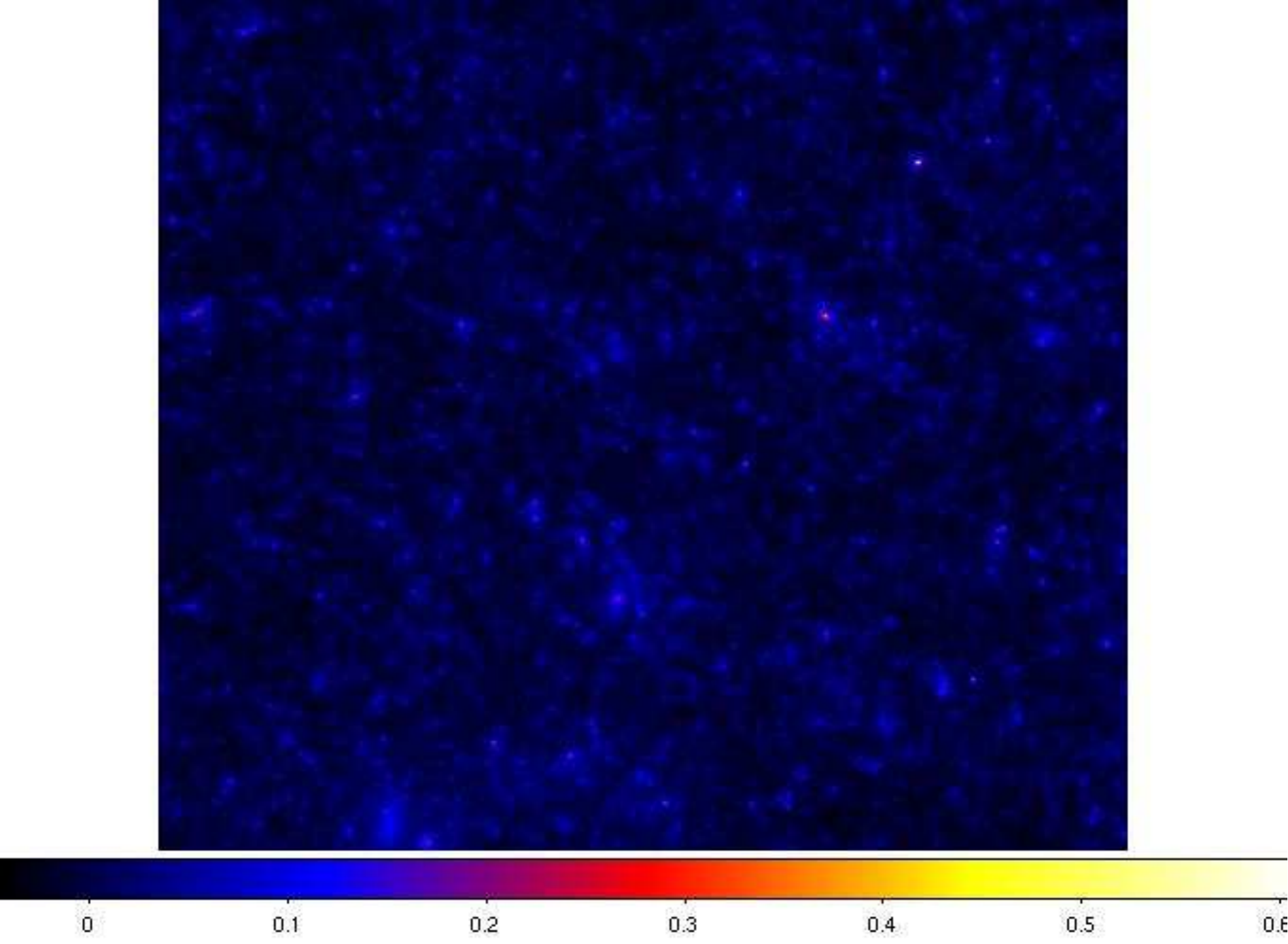}
   \includegraphics[width=7.5cm, angle=0]{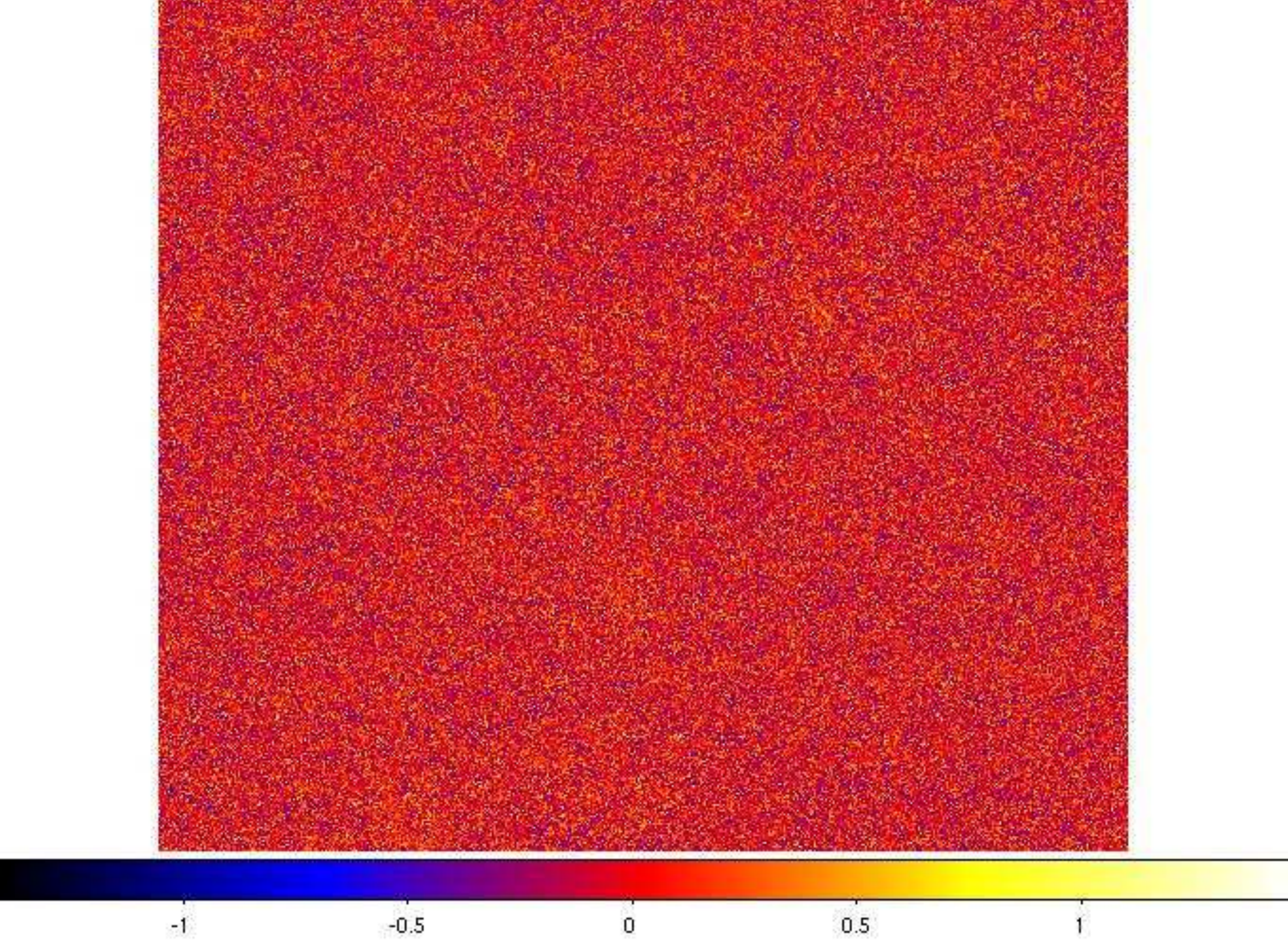}
   \caption{ The convergence maps of $3\times3\hbox{ deg}^2$.
The left panel is the noise-free map, and the right panel is the
noisy convergence map with $n_{g}=30\hbox{ arcmin}^{-2}$.}
   \label{Fig1}
   \end{figure}

\subsection{Statistical properties of residual noise}
\label{sect:noise}

Post-processing procedures can reduce noise effectively. However,
certain levels of residual noise inevitably remain. It is thus
important to understand the statistical properties of the residual
noise so that we can quantify their effects on weak-lensing
cosmological studies properly. For that, we first in this subsection
consider pure noise maps without including weak-lensing signals.
After applying Gaussian smoothing and MRLens, respectively, we
compare the residual noise-peak statistics in the two cases. This is
highly relevant to cosmological applications of weak-lensing cluster
statistics, in which, high peaks in convergence maps are thought to
be related to clusters of galaxies and their abundances contain
important cosmological information. The existence of residual noise
can generate false peaks in convergence maps, which in turn can
contaminate the weak-lensing peak statistics significantly.

With Eq.~(30), we generate a $3\times 3\hbox{ deg}^2$ noise map
containing $1024\times 1024$ pixels with $\theta_{pix}=0.176\hbox{
arcmin}$ and the corresponding $\sigma_{pix}=0.22$ for $n_g=30\hbox{
arcmin}^{-2}$ and $\sigma_{pix}=0.12$ for $n_g=100\hbox{ arcmin}^{-2}$. 
For Gaussian smoothing, we take $\theta_G=1\hbox{ arcmin}$.
For MRLens, we take $\alpha_0=0.01$. In a smoothed map, a
positive (maximum)/negative (minimum) peak position is located if
its value is above/below those of its eight neighboring pixels
(e.g., Jain \& Van Waerbeke 2000; Miyazaki et al. 2002).

Figure~2 shows the probability distribution function (PDF) of peaks in the residual noise field
for the two cases, respectively, with the left
for the Gaussian smoothing and the right for the MRLens. In each panel,
the solid and dashed lines correspond to the results with $n_g=30\hbox{ arcmin}^{-2}$
and $n_g=100\hbox{ arcmin}^{-2}$, respectively. The bin
size is $\Delta\kappa=0.005$. Both the positive and the negative
peaks are counted in. Two distinctly different distributions are
seen. For the Gaussian smoothing case, the peak number distribution has a
double-peak behavior at $\kappa/\sigma_0\sim \pm 1$, in good agreement with that expected
for a Gaussian random field (Bond \& Efstathiou 1987; Van Waerbeke
2000). The rms of the residual noise in this case is
$\sigma_0\approx 0.016$ for $n_{g}=30\hbox{ arcmin}^{-2}$ and 
$\sigma_0\approx 0.009$ for $n_{g}=100\hbox{ arcmin}^{-2}$, in excellent agreement with the theoretical
value $0.015$ for $n_{g}=30\hbox{ arcmin}^{-2}$ and $\sigma_0\approx 0.008$ for $n_{g}=100\hbox{ arcmin}^{-2}$
calculated from Eq.~(15). Considering positive peaks that are relevant for weak-lensing analyses, the
noise peaks with $\kappa/\sigma_0\sim 1$, rather than with
$\kappa/\sigma_0=0$, have the highest occurrence probability. Such a
property of noise can cause statistically a positive shift for the
peak height of a cluster measured from noisy convergence maps. The
shift depends on the density profile of the cluster. This
noise-induced shift can bias the cluster mass estimation from
weak-lensing observations. On the other hand, it can increase the
weak-lensing detectability of clusters, and thus affect the
corresponding cosmological studies significantly (Fan et al. 2010).

For the MRLens case, the residual noise after restoration treatment
is low with $\sigma_0\approx 0.0029$ for $n_{g}=30\hbox{ arcmin}^{-2}$
and $\sigma_0\approx 0.0016$ for $n_{g}=100\hbox{ arcmin}^{-2}$, much less than those of the
Gaussian smoothing. However, the noise statistics is highly
non-Gaussian, which results significant complications in quantifying the
noise effect on weak-lensing signals. The number distribution of
noise peaks is narrowly concentrated around $\kappa=0$. Thus unlike
the Gaussian smoothing, it seems that we do not expect a systematic
shift due to noise in weak-lensing cluster peak measurement. It
should be noted, however, in MRLens, the noise filtering involves
restoration procedures based on NOISE-MSE of Eq.~(29). The results
depend on the noise properties [the second term in Eq.~(29)] as well
as on the properties of signals we would like to detect [the first
term in Eq.~(29)]. The higher the original noise is, the larger the
fraction of the wavelet coefficients that are suppressed. In such a
treatment, the signals are changed depending on the original noise
level and their own properties. Therefore considering the
convergence peak for a cluster, the results after MRLens restoration
in the cases with and without noise are different. In this sense,
the existence of noise also induces a systematic bias for the peak
value of a cluster, though for a reason different from and much more
complicated than that of the Gaussian smoothing case. The
quantitative modeling of such a bias for MRLens needs to be further
explored.

\begin{figure}
\centering
\includegraphics[width=10.0cm]{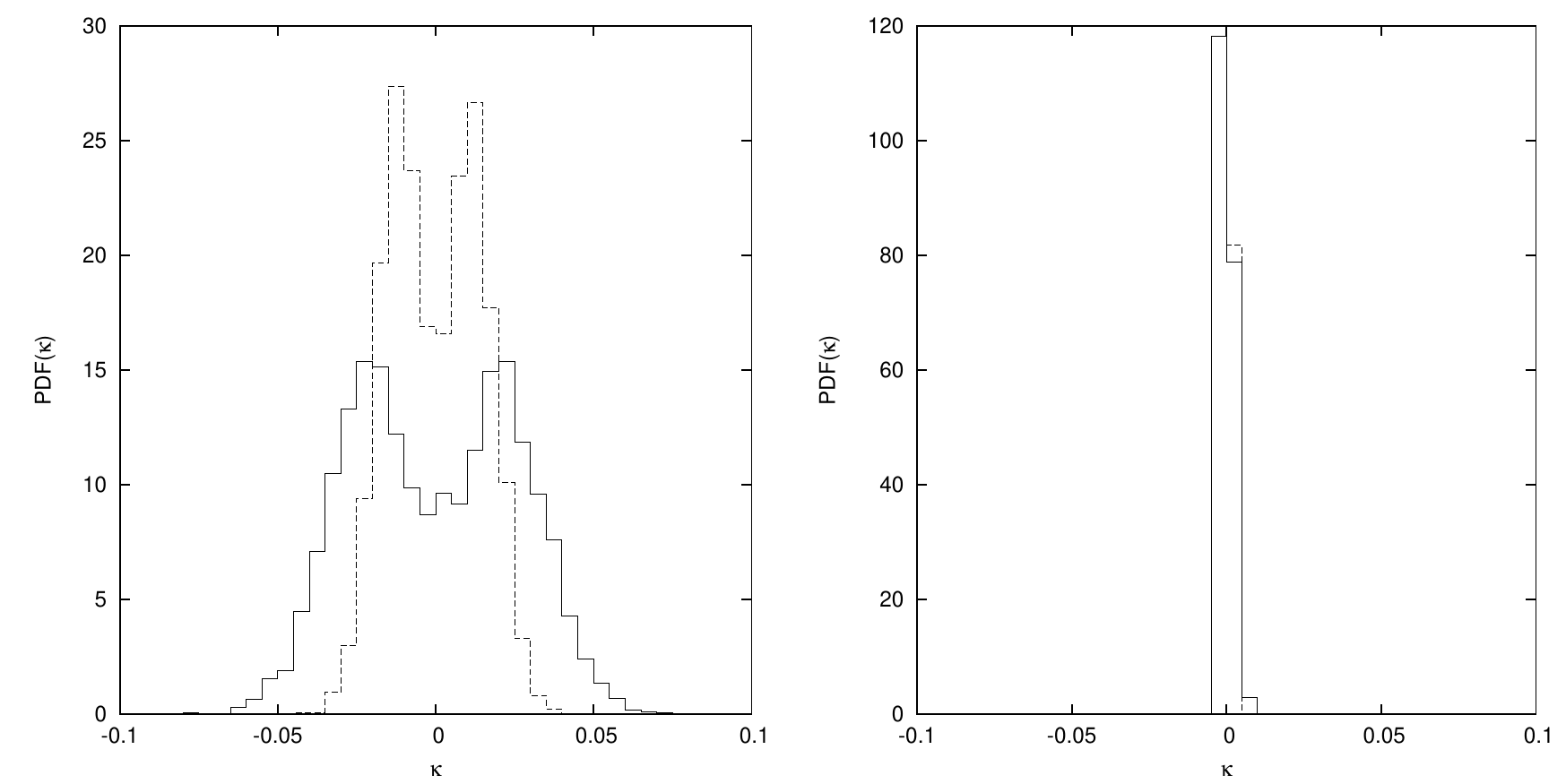}
\caption{The probability distribution functions of peaks in pure
noise maps. The solid and dashed lines are for $n_{g}=30\hbox{ arcmin}^{-2}$
and $n_{g}=100\hbox{ arcmin}^{-2}$, respectively.
The left panel is for the Gaussian smoothing with $\theta_{G}=1\hbox{ arcmin}$, and the right panel is
for the MRLens result with $\alpha_0=0.01$.} \protect\label{fig2}
\end{figure}

For MRLens, the $\alpha_0$ parameter in FDR affects the
classification of significant and non-significant wavelet
coefficients. A smaller $\alpha_0$ results a more stringent criteria
for the definition of a significant wavelet coefficient, and thus
stronger suppressions of noise. To test the $\alpha_0$-dependence,
we vary its value to obtain different restoration results for pure
noise maps. In Table~1, the rms of the residual noise for different
$\alpha_0$ and different $n_g$ are shown. With the increase of
$n_g$, the original noise level decreases with $(n_g)^{-1/2}$. It is
noted that after MRLens treatment, the rms of the residual noise
also approximately follows $\sigma_0\propto (n_g)^{-1/2}$. For the
$\alpha_0$-dependence, as expected, the residual noise decreases
with the decrease of $\alpha_0$. However, this dependence is rather
weak. Changing $\alpha_0$ from $0.1$ to $0.01$ only decreases
$\sigma_0$ by $\sim 20\%$.

\begin{table}
\centering \caption{Standard deviation of the reconstruction error
with MRLens}
 \begin{tabular}{ccccc}
  \hline
  \hline
$\alpha_0$ & $\sigma_{0}$($n_{g}=15$) & $\sigma_{0}$($n_{g}=30$) &  $\sigma_{0}$($n_{g}=50$)  &
$\sigma_{0}$($n_{g}=100$)  \\
  \hline
0.001   & 0.0038 & 0.0026  & 0.0021   & 0.0015 \\
0.01    & 0.0041 & 0.0029  & 0.0023   & 0.0016 \\
0.02    & 0.0041 & 0.0030  & 0.0023   & 0.0016 \\
0.04   & 0.0044 & 0.0032  & 0.0024   & 0.0017 \\
0.06   & 0.0045 & 0.0033  & 0.0026   & 0.0019 \\
0.08   & 0.0047 & 0.0033  & 0.0027   & 0.0020 \\
0.1    & 0.0050 & 0.0035  & 0.0027   & 0.0021 \\
0.2    & 0.0051 & 0.0036  & 0.0029   & 0.0023 \\
\hline
\end{tabular}
\label{tab:sde}
\end{table}

\subsection{Peak statistics in noisy convergence maps}

Now we consider peak statistics of noisy convergence maps. Figure~3
shows the post-processed maps of the right panel of Figure~1 with
Gaussian smoothing for $\theta_G=1\hbox{ arcmin}$ (upper) and with
MRLens for $\alpha_0=0.01$ (lower), respectively. The left panels are
for $n_{g}=30\hbox{ arcmin}^{-2}$ and the right panels are for $n_{g}=100\hbox{ arcmin}^{-2}$.

\begin{figure}[h!!!]
 \includegraphics[width=7.5cm, angle=0]{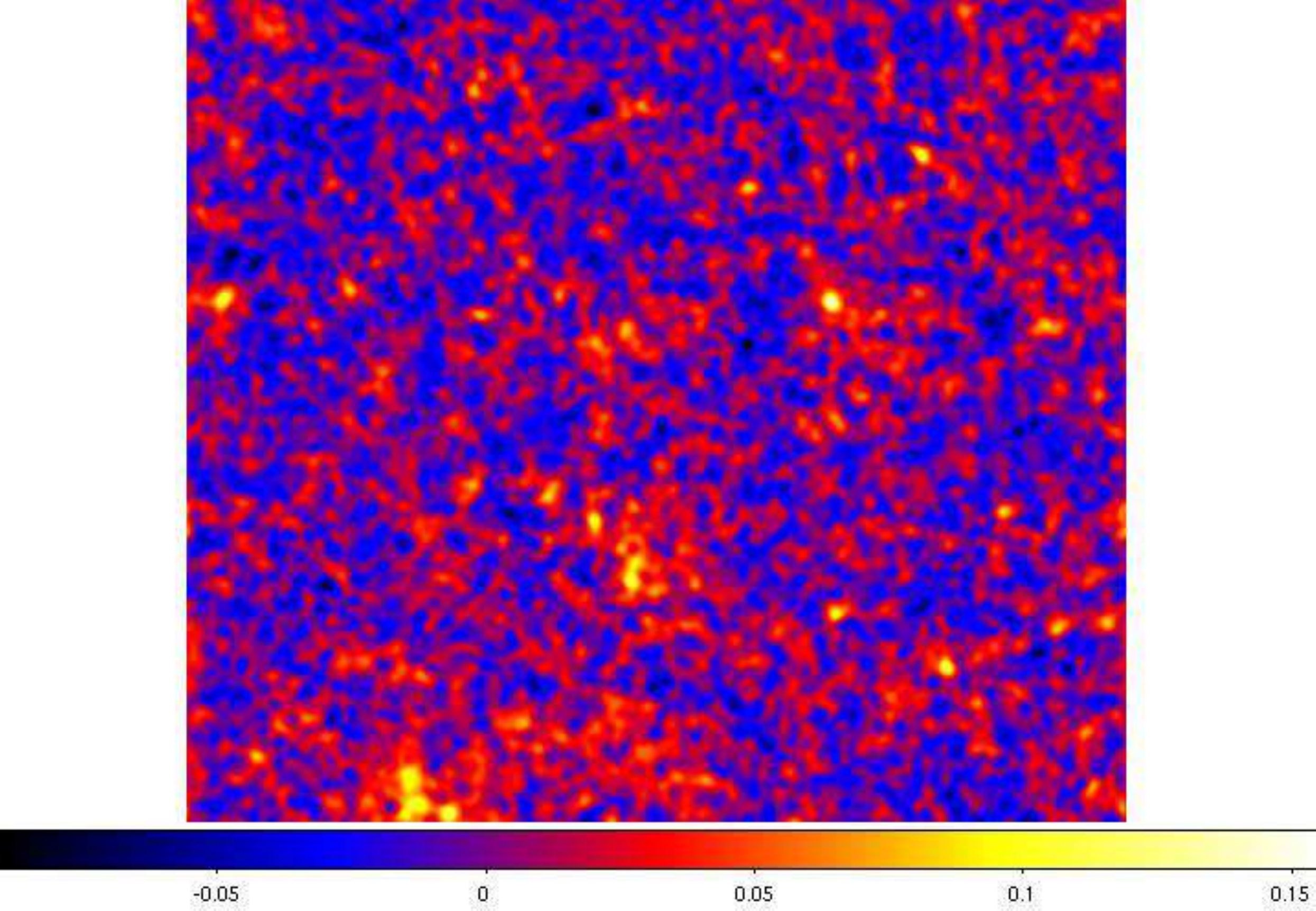} 
 \includegraphics[width=7.5cm, angle=0]{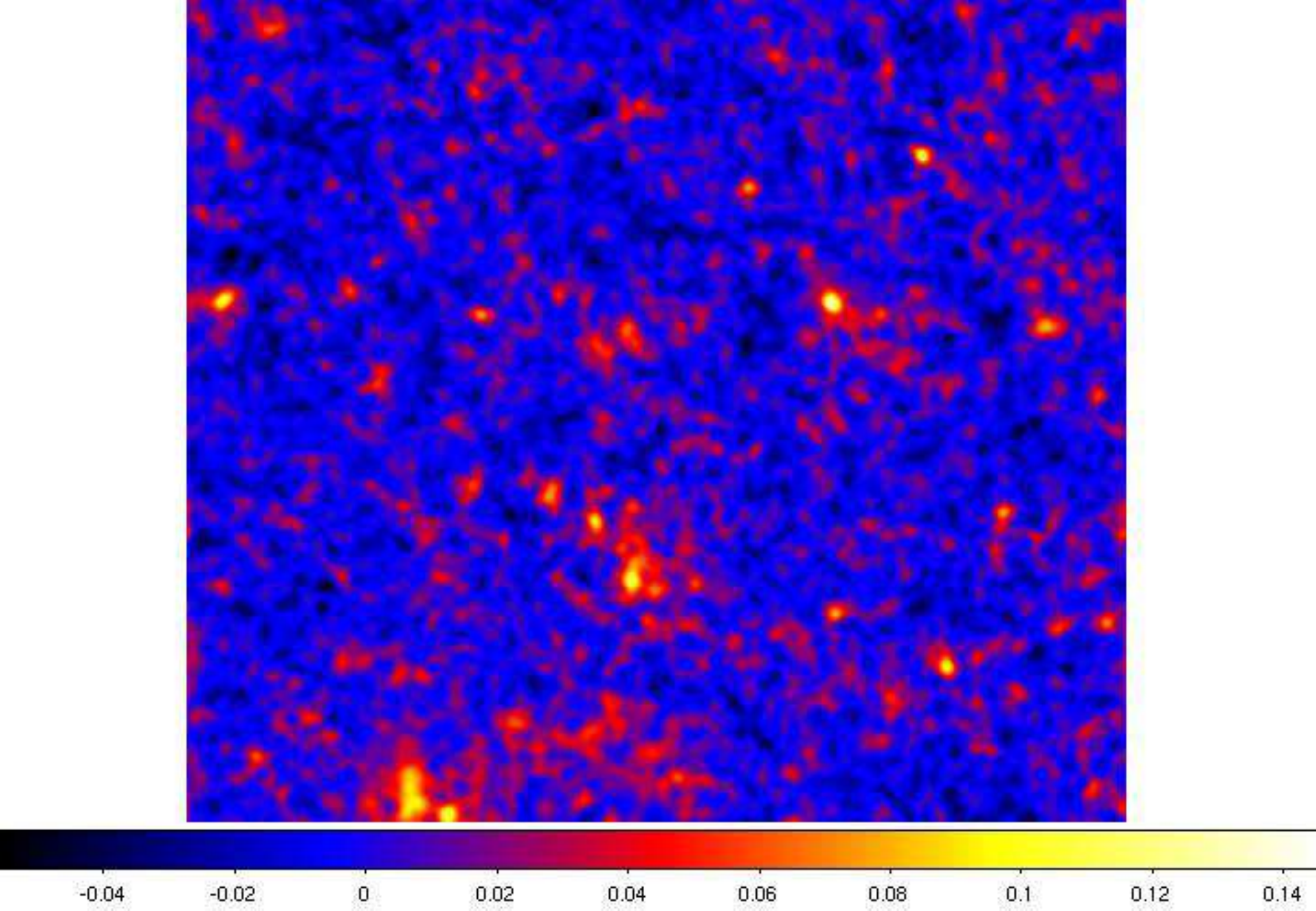}
 \includegraphics[width=7.5cm, angle=0]{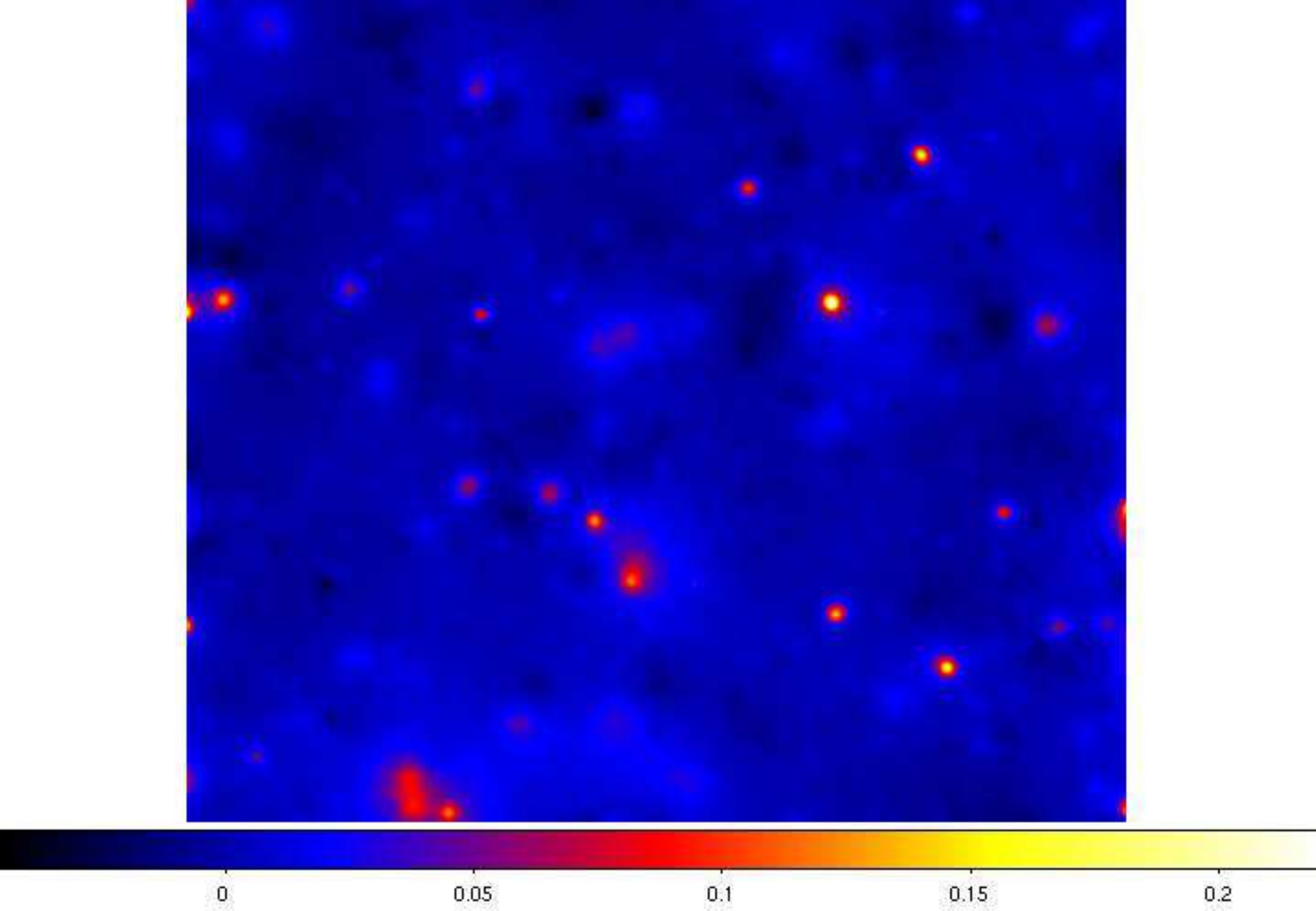}
 \includegraphics[width=7.5cm, angle=0]{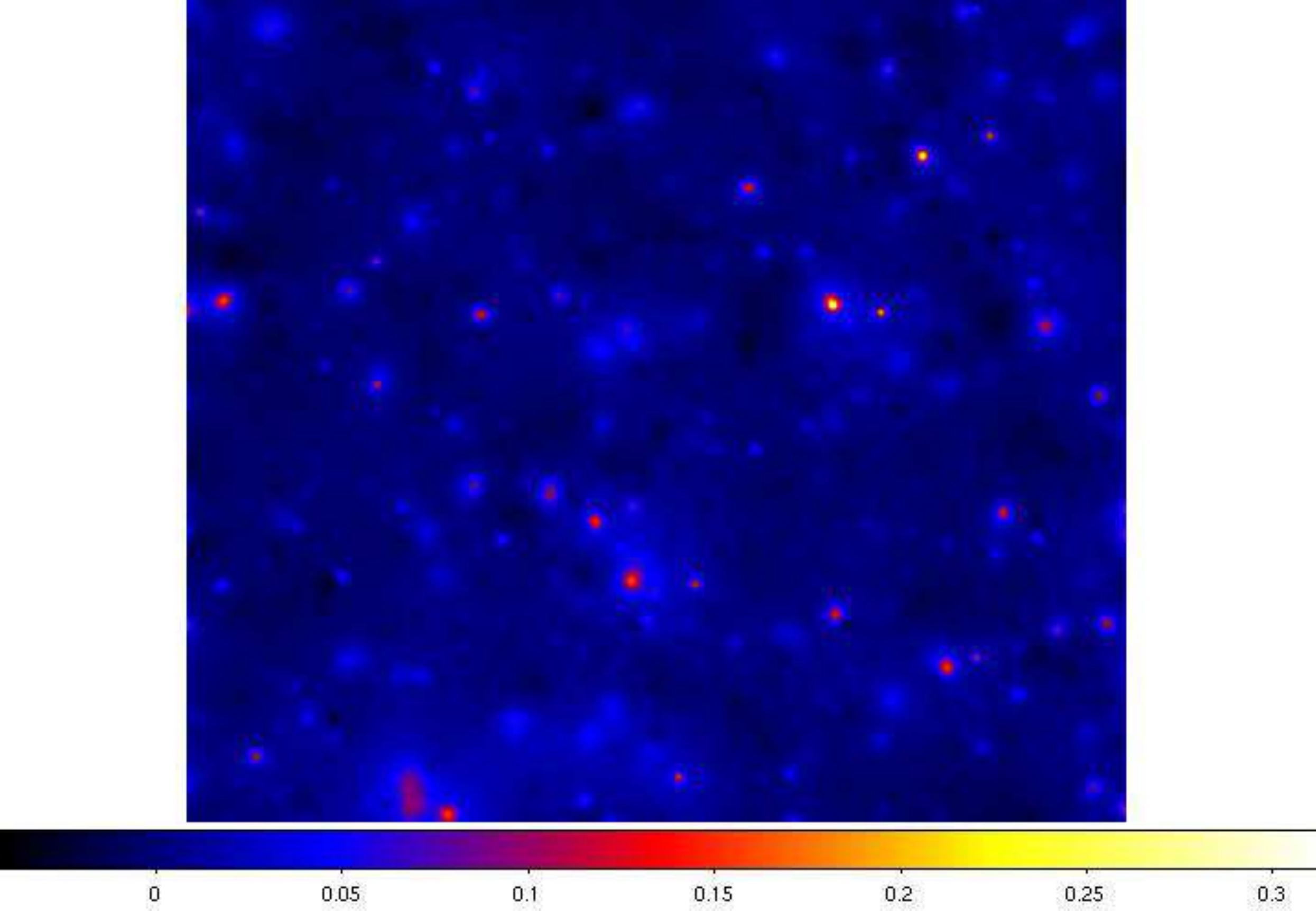}
\caption{ Noisy convergence maps of $3\times3 \hbox{ deg}^2$. The upper and lower panels are for
the Gaussian smoothing with $\theta_{G}=1\hbox{ arcmin}$ and MRLens with $\alpha_0=0.01$, respectively. The left and right panels are for
$n_{g}=30\hbox{ arcmin}^{-2}$ and $n_{g}=100\hbox{ arcmin}^{-2}$, respectively.}
   \label{Fig3}
   \end{figure}

Comparing to the maps in Figure~1, we
see that the post-processing procedures can indeed filter out much
of the noise so that the real structures in the large-scale mass
distribution can be detected. For $n_{g}=30\hbox{ arcmin}^{-2}$, the MRLens map looks very smooth with
only very massive structures left. On the other hand, in the
Gaussian smoothing case, small structures can also be seen. However,
it contains many more noise peaks than that of the MRLens case.
For $n_{g}=100\hbox{ arcmin}^{-2}$, the map is smoother for the Gaussian smoothing case 
than that with $n_{g}=30\hbox{ arcmin}^{-2}$. 
The MRLens map, however, appears lumpier for the lower noise case. 
Such opposite trends seen in the Gaussian smoothing and in MRLens 
reflect clearly the different underlying filtering mechanisms between the two smoothing schemes.
For the Gaussian smoothing, the filtering is mainly performed
through an averaging procedure. Given a smoothing scale, the peak signals of
real clusters are more or less similar regardless the noise level. Meanwhile, the noise peaks 
with relatively high $\kappa$ values are significantly reduced if the noise level is lowered. 
Thus the smoother appearance of the upper right panel
is mainly due to the less number of high noise peaks than that of the upper left panel. 
For MRLens, it involves a restoration procedure that depends on the original noise
level. The smaller the original noise is, the lower the fraction is for
the wavelet coefficients to be suppressed. It is important to note that 
the suppression leads to the removal of both noise peaks and true peaks
of relatively low amplitudes. Thus the lumpier structures seen in the lower
right panel is largely attributed to the lower level of removal of real structures
than that of the lower left panel.

In Figure~4 and Figure~5, we show the probability distribution
function of peaks for Gaussian smoothing and for MRLens,
respectively. The results for each case are obtained by averaging
over $16$ simulated maps with noise added.

\begin{figure}
\centering
\includegraphics[width=10.0cm]{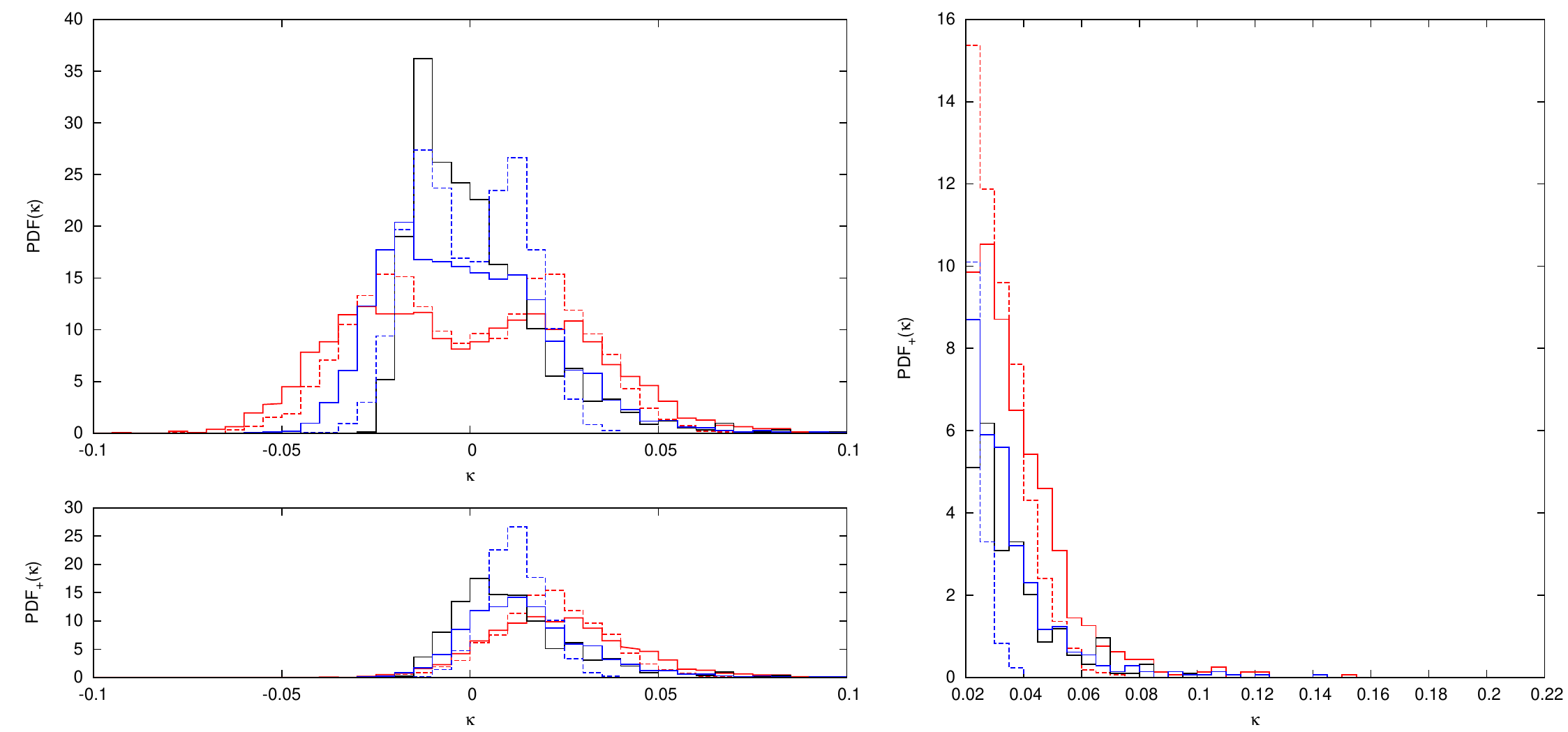}
\caption{The probability distribution function of convergence peaks
for the case of Gaussian smoothing with $\theta_G=1\hbox{ arcmin}$. 
The black solid line is for the result
of the noise-free convergence peaks, the red dashed and red solid lines are for the pure
noise peaks and noisy convergence peaks with $n_g=30\hbox{ arcmin}^{-2}$, respectively,
and the blue dashed and blue solid lines are for the pure noise peaks
and noisy convergence peaks with $n_g=100\hbox{ arcmin}^{-2}$, respectively.
The left upper panel includes both maximum and minimum
peaks, and the left lower panel shows the distribution function for
maximum peaks only. The right panel is the zoom-in version of the left lower panel.} \protect\label{fig4}
\end{figure}

\begin{figure}
\centering
\includegraphics[width=10.0cm]{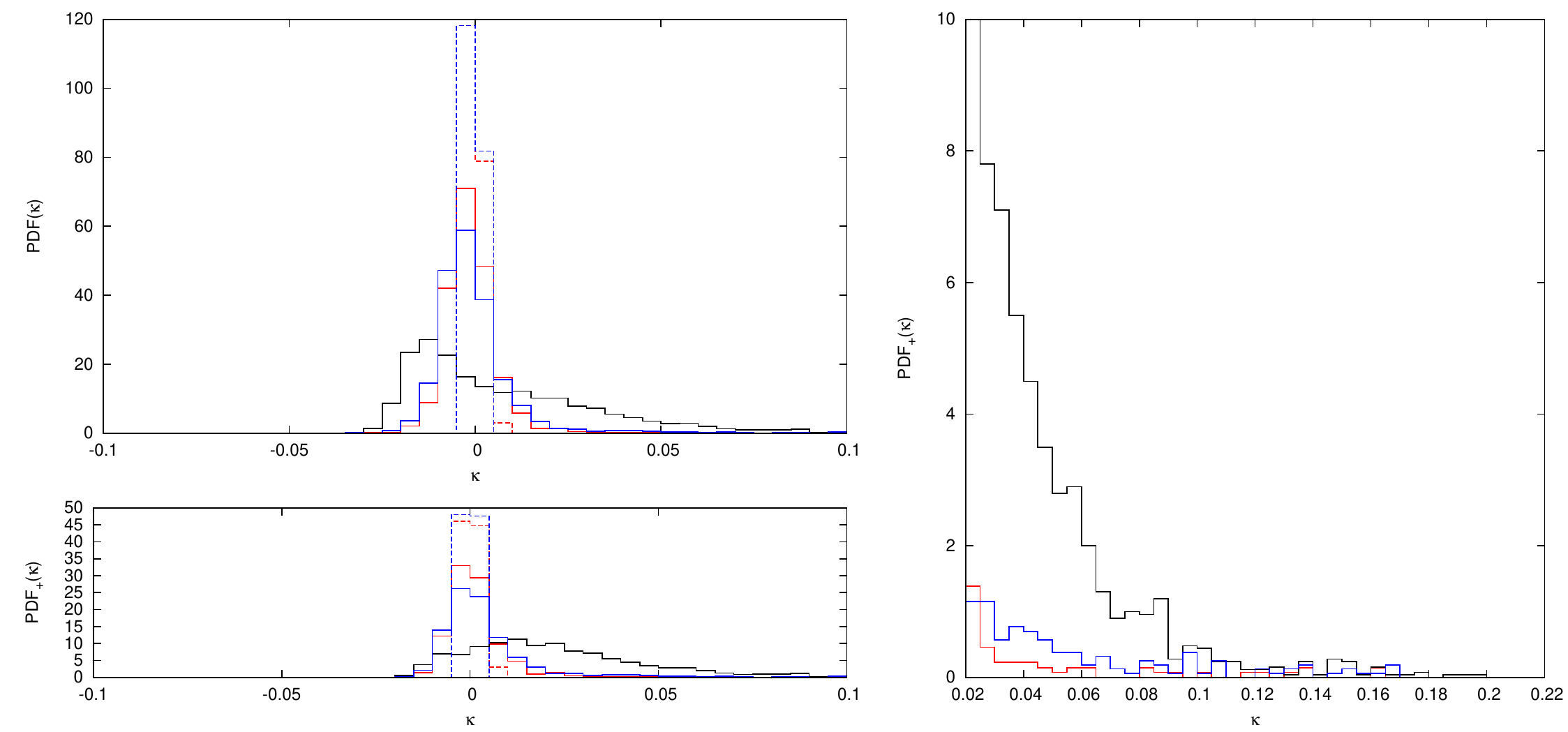}
\caption{Same as Figure~4 but for the case of MRLens with $\alpha_0=0.01$. }
\protect\label{fig5}
\end{figure}

For the Gaussian smoothing results in Figure~4, the black, red dashed, red solid, blue dashed, 
and blue solid lines are for the results of noise free peaks, pure noise peaks with $n_g=30\hbox{ arcmin}^{-2}$, noisy convergence
peaks with $n_g=30\hbox{ arcmin}^{-2}$, pure noise peaks with $n_g=100\hbox{ arcmin}^{-2}$, noisy convergence
peaks with $n_g=100\hbox{ arcmin}^{-2}$, respectively.
We can see that in the Gaussian smoothing cases, the noise peaks dominate over the 
real peaks at $\kappa<3\sigma_0$. At larger $\kappa>3\sigma_0$, real peaks can be detected with
high efficiencies. Comparing the blue solid line with the red solid line, we see that by reducing the noise level
from $\sigma_0\sim 0.015$ to $\sigma_0\sim 0.008$, we effectively reduce the number of noise peaks with $\kappa>0.025$, 
and thus increase the real peak detection efficiencies significantly.

In Figure~5 for the MRLens results, the line styles are the same as those in Figure~4.
Different from that in the Gaussian smoothing cases, here the noise peaks (red and blue dashed lines) contribute little
to the total number of peaks with $\kappa>0.02$ in comparison with the real peaks (black solid line). However, 
the suppression process in the MRLens treatment mistakenly removes a large number of real peaks
with $\kappa<0.1$. Thus we expect a high efficiency but a low completeness in weak-lensing peak detections 
after MRLens filtering. Reducing the original noise level 
by increasing $n_g$ form $30\hbox{ arcmin}^{-2}$ to $100\hbox{ arcmin}^{-2}$
leads to a less suppression effect. 
Therefore more peaks with $\kappa<0.1$ are kept and the completeness of peak detections increases considerably.  

In the next subsection, we investigate and compare explicitly the efficiency and completeness of 
weak-lensing cluster detections in the two smoothing treatments. 

\subsection{Efficiency and completeness of weak-lensing cluster detection}
\label{sect:ef}

The existence of noise from intrinsic ellipticities of source
galaxies results false peaks in convergence maps, and thus lowers
considerably the efficiency of weak-lensing cluster detections.
Increasing the detection threshold can increase the efficiency,
however at the expense of completeness. In this section, we compare
the weak-lensing cluster detection with Gaussian smoothing and with
the MRLens, respectively. Following Hamana et al. (2004), we define
the efficiency $f_{e}$ and completeness $f_{c}$ of cluster detection
with respect to the number of clusters (dark matter halos) above a
certain mass threshold. Specifically, we have
\begin{equation}
f_{e}=\frac{N_{iii}}{N_{i}},
\end{equation}
\begin{equation}
f_{c}=\frac{N_{iii}}{N_{ii}},
\end{equation}
where $N_{i}$ denotes the number of convergence peaks with their
heights above a detection threshold, $N_{ii}$ represents the number
of dark matter halo with mass above a certain mass threshold, and
$N_{iii}$ is the number of peaks that have correspondences with dark
matter halos among $N_{ii}$. A peak is defined to be associated with
its nearest dark matter halo if the location of the peak is within a
radius of $12$ pixels (corresponds to $2.11\hbox{ arcmin}$) around
the halo. If there are two or more peaks associated with a same
halo, the highest peak is defined to have the correspondence with
the halo.

\begin{figure}
\centering
\includegraphics[width=6.5cm]{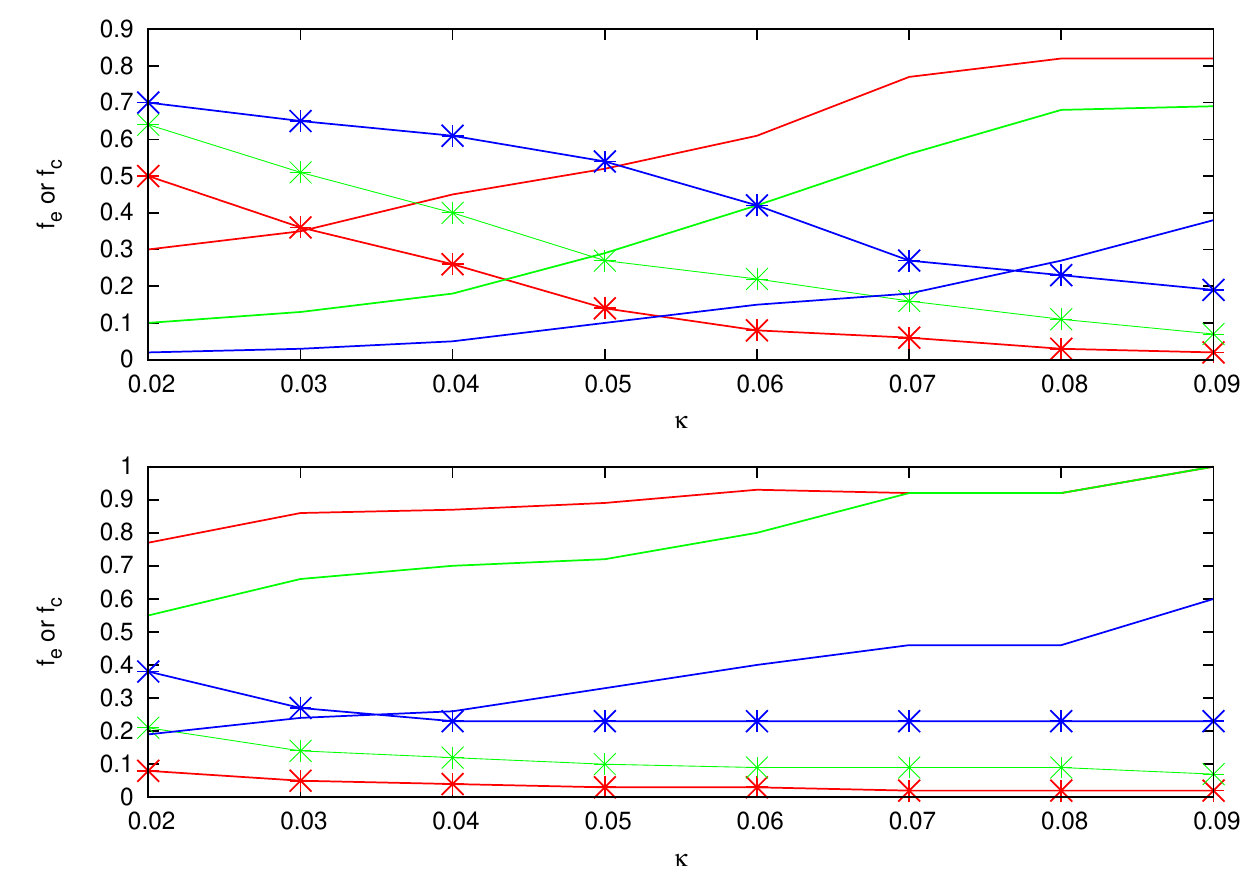}
\includegraphics[width=6.5cm]{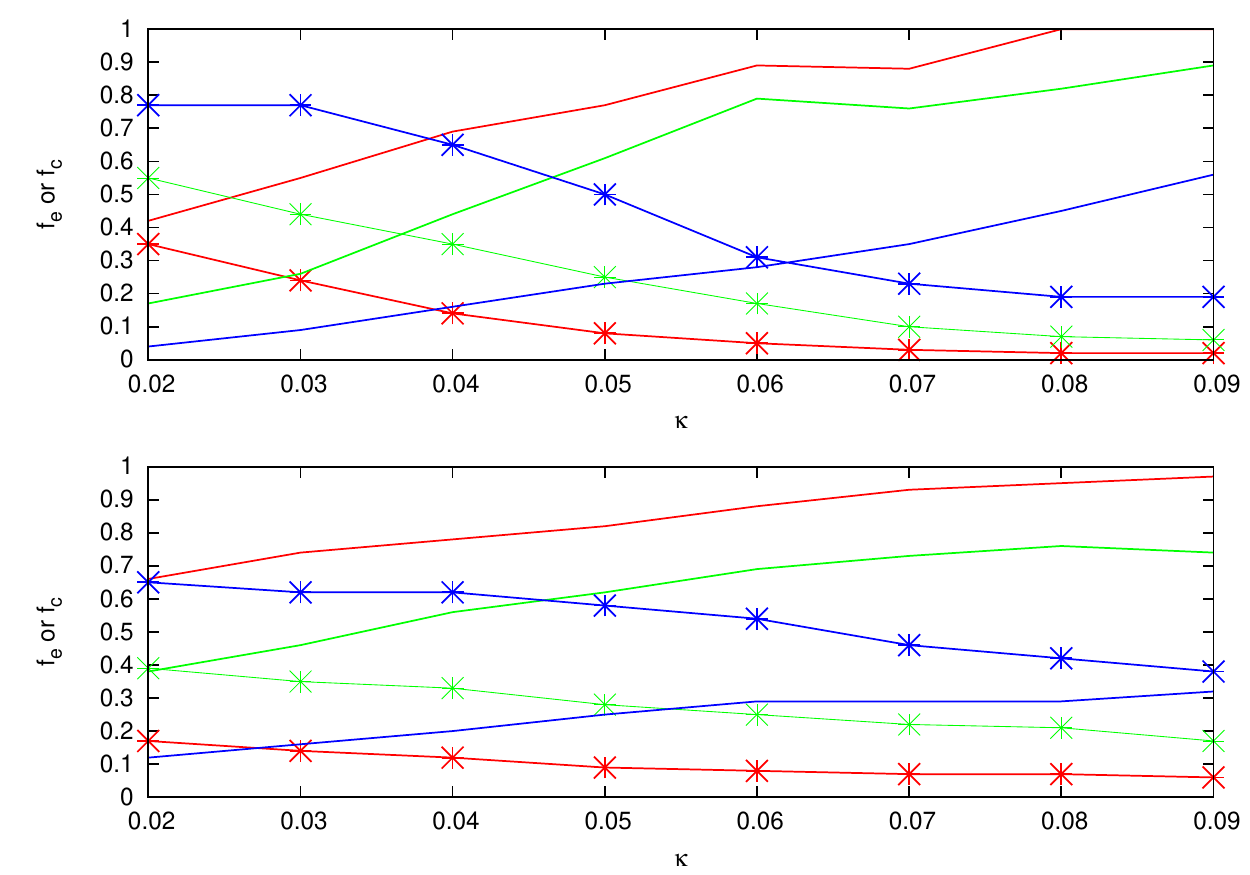}
\caption{The efficiency $f_e$ and completeness $f_c$ as functions of
the peak detection threshold $\kappa$. The left and right panels are for $n_g=30\hbox{
arcmin}^{-2}$ and $n_g=100\hbox{ arcmin}^{-2}$, respectively. 
The upper two panels are for the Gaussian smoothing and the
lower two panels are for MRLens. The lines with symbols are for the
completeness, and the lines without symbols are for the efficiency.
The red, green and blue lines are for the results of halos with $M
>5\times10^{13} M_{\odot}$, $M >1\times10^{14} M_{\odot}$, and $M
>2\times10^{14} M_{\odot}$, respectively.} \protect\label{fig6}
\end{figure}

Figure~6 shows the results of $f_e$ and $f_c$ for Gaussian smoothing (upper panels)
and MRLens (lower panels). The left panels are for $n_g=30\hbox{ arcmin}^{-2}$, 
and the right panels are for $n_g=100\hbox{ arcmin}^{-2}$. 
In each panel, the red, green and blue lines are for halos with mass $M>5\times
10^{13}M_{\odot}$, $M>1\times 10^{14}M_{\odot}$, and $M>2\times
10^{14}M_{\odot}$, respectively. The lines with and without symbols
are, respectively, for the results of completeness and efficiency.
The horizontal axis in each panel is the peak detection threshold
$\kappa$.

We first analyze the Gaussian smoothing cases. As we discuss previously,
such a smoothing process reserves more or less all the real peaks
with scales above the smoothing scale. At mean time, the number of noise peaks
is large at $\kappa<3\sigma_0$. Thus a high completeness and a low
efficiency are expected when the peak detection threshold is low. 
For $n_g=30\hbox{ arcmin}^{-2}$ (upper left), we have $\sigma_0\sim 0.015$. At the detection 
threshold $\kappa=0.02\sim 1.3\sigma_0$, we have the completeness $f_c\sim 50\%, 65\%$ and $70\%$ for 
$M>5\times 10^{13}\hbox{ M}_{\odot}$, $1\times 10^{14}\hbox{ M}_{\odot}$
and $2\times 10^{14}\hbox{ M}_{\odot}$, respectively. The corresponding efficiencies
are $30\%, 10\%$ and $2\%$. When the detection threshold $\kappa>3\sigma_0$,
the number of noise peaks drops significantly, leading to a large increase
in the detection efficiency. On the other hand, a considerable fraction of halos
are missed due to the high detection threshold, resulting a decrease
in the completeness. Specifically, at $\kappa=0.045\sim 3\sigma_0$,
the completeness $f_c\sim 20\%, 35\%$ and $60\%$, and the efficiency
$f_e\sim 50\%, 25\%$ and $10\%$, for $M>5\times 10^{13}\hbox{ M}_{\odot}$, $1\times 10^{14}\hbox{ M}_{\odot}$
and $2\times 10^{14}\hbox{ M}_{\odot}$, respectively.
With the increase of $n_g$ to $n_g=100\hbox{ arcmin}^{-2}$ (upper right), the noise level
$\sigma_0$ decreases by a factor of $\sqrt{100/30}$ to $\sigma_0\sim 0.008$. 
Thus $3\sigma_0$ corresponds to $\kappa\sim 0.025$. At this detection threshold,
the number of noise peaks is smaller and correspondingly the efficiency is higher
than those with $n_g=30\hbox{ arcmin}^{-2}$. On the other hand, the number of real peaks
does not change much as the noise level decreases, and
thus the completeness is similar to that of $n_g=30\hbox{ arcmin}^{-2}$.  
Quantitatively, at the threshold $\kappa=0.025$, the efficiency $f_e\sim 50\%, 
20\%$ and $8\%$, in comparison with $f_e\sim 35\%, 12\%$ and $3\%$ in the case of $n_g=30\hbox{ arcmin}^{-2}$,
for $M>5\times 10^{13}\hbox{ M}_{\odot}$, $1\times 10^{14}\hbox{ M}_{\odot}$
and $2\times 10^{14}\hbox{ M}_{\odot}$, respectively. For the completeness, 
we have $f_c\sim 30\%, 50\%$ and $80\%$ for $n_g=100\hbox{ arcmin}^{-2}$.
For $n_g=30\hbox{ arcmin}^{-2}$, $f_c\sim 45\%, 60\%$ and $70\%$. While being similar, 
$f_c$ decreases somewhat for $M>5\times 10^{13}\hbox{ M}_{\odot}$ and $1\times 10^{14}\hbox{ M}_{\odot}$ 
with the decrease of noise level. This is in accordance with the analyses of Fan et al. (2010)
where they find that the existence of noise generates a systematic shift for the real peaks toward
higher amplitudes. The shift depends on the density profile of dark matter halos associated with the 
real peaks, and can be as high as $\sim 1\sigma_0$ for NFW halos with low concentrations. 
In terms of $\kappa$ values, the shift is larger for larger $\sigma_0$. 
Thus, in the case of $n_g=30\hbox{ arcmin}^{-2}$, the relatively large $\sigma_0$ leads to 
a large shift of the real peak heights and consequently a larger number of real peaks above
the detection threshold than that in the case of $n_g=100\hbox{ arcmin}^{-2}$. 

For MRLens, with $n_g=30\hbox{ arcmin}^{-2}$ (lower left), the completeness of the weak-lensing cluster detection is
very low, and $f_c\sim 10\%, 20\%$ and $40\%$ at the threshold $\kappa=0.02$, in comparison with 
$f_c\sim 50\%, 65\%$ and $70\%$ in the corresponding Gaussian smoothing case. This is because the 
suppression of the wavelet coefficients aiming to reduce noise removes a large fraction of 
real peaks in the range of $\kappa<0.1$ as seen from Figure~5. The total number of peaks
in $3\times 3\hbox{ deg}^2$ with $\kappa\ge 0.02$ is only $\sim 53$, while the total number of halos 
in the area with $M>5\times 10^{13}\hbox{ M}_{\odot}$ is $\sim 530$. Thus although the efficiency in MRLens
here is rather high ($\sim 80\%$ for $M>5\times 10^{13}\hbox{ M}_{\odot}$), the very few number of 
detected halos makes the MRLens method be disadvantageous in comparing with that of simple Gaussian smoothing
method. For $n_g=100\hbox{ arcmin}^{-2}$, the noise level is lower and thus the removal effect is
less significant than the case of $n_g=30\hbox{ arcmin}^{-2}$. Consequently, the completeness 
increases considerably with $f_c\sim 20\%, 40\%$ and $65\%$. Meanwhile, the efficiency decreases
somewhat. In this low noise case, the differences between the MRLens and Gaussian smoothing
in terms of the completeness and efficiency are less than those of high noise case. 
But still, the completeness is lower for MRLens, especially considering relatively low mass halos
with $M>5\times 10^{13}\hbox{ M}_{\odot}$. 

To further demonstrate the differences between the Gaussian smoothing and the MRLens, in Figure~7, 
we show the peak-halo correspondences explicitly in
$z-M$ plane for one of our $3\times 3\hbox{ deg}^{2}$ simulation maps, 
where $M$ is the halo mass in unit of $10^{13}M_{\odot}$ and $z$ is the halo
redshift from simulations. The halos are the ones located in the solid angle of
$3\times 3\hbox{ deg}^{2}$ in the considered direction and their redshift and mass 
are taken directly from the halo catalogs constructed by White and Vale (2004). 
The upper and lower panels are for the Gaussian smoothing and the MRLens, respectively. 
The left and right panels correspond to $n_g=30\hbox{ arcmin}^{-2}$ and $n_g=100\hbox{ arcmin}^{-2}$, 
respectively. In each panel, the `+' symbols denote
the dark matter halos identified in simulations with $M\ge 5\times
10^{13} M_{\odot}$ and in the redshift range of $0\le z\le 2$. There are very few halos
extending to redshift beyond $z=2$. The
green squares represent those halos that have corresponding
convergence peaks with $\kappa\ge 0.02$. The differences between the two filtering methods are 
strikingly seen. For MRLens with $n_g=30\hbox{ arcmin}^{-2}$ (lower left), a majority of halos with $M<10^{14}\hbox{ M}_{\odot}$
or with $z>0.8$ are missed in weak-lensing detections, consistent with its extremely low
completeness shown in Figure~6. Lowering the noise level by increasing $n_g$ to $n_g=100\hbox{ arcmin}^{-2}$
increases the number of halos with associated peaks by nearly a factor of $2$ (lower right). But the number
is still much less than that in the Gaussian smoothing case. Therefore in studies aiming to 
detect a large number of clusters from blind surveys and subsequent cosmological applications,
the Gaussian smoothing method is clearly much better than the MRLens. In addition, 
the noise field after a Gaussian smoothing with $\theta_G\sim 1\hbox{ arcmin}$ is approximately
Gaussian in statistics, and thus its effects on weak-lensing cluster detection can be modeled much easier  
than the case of MRLens where the left-over noise is statistically highly non-Gaussian (Fan et al. 2010). 

\begin{figure}[h!!!]
   \includegraphics[width=7.5cm, angle=0]{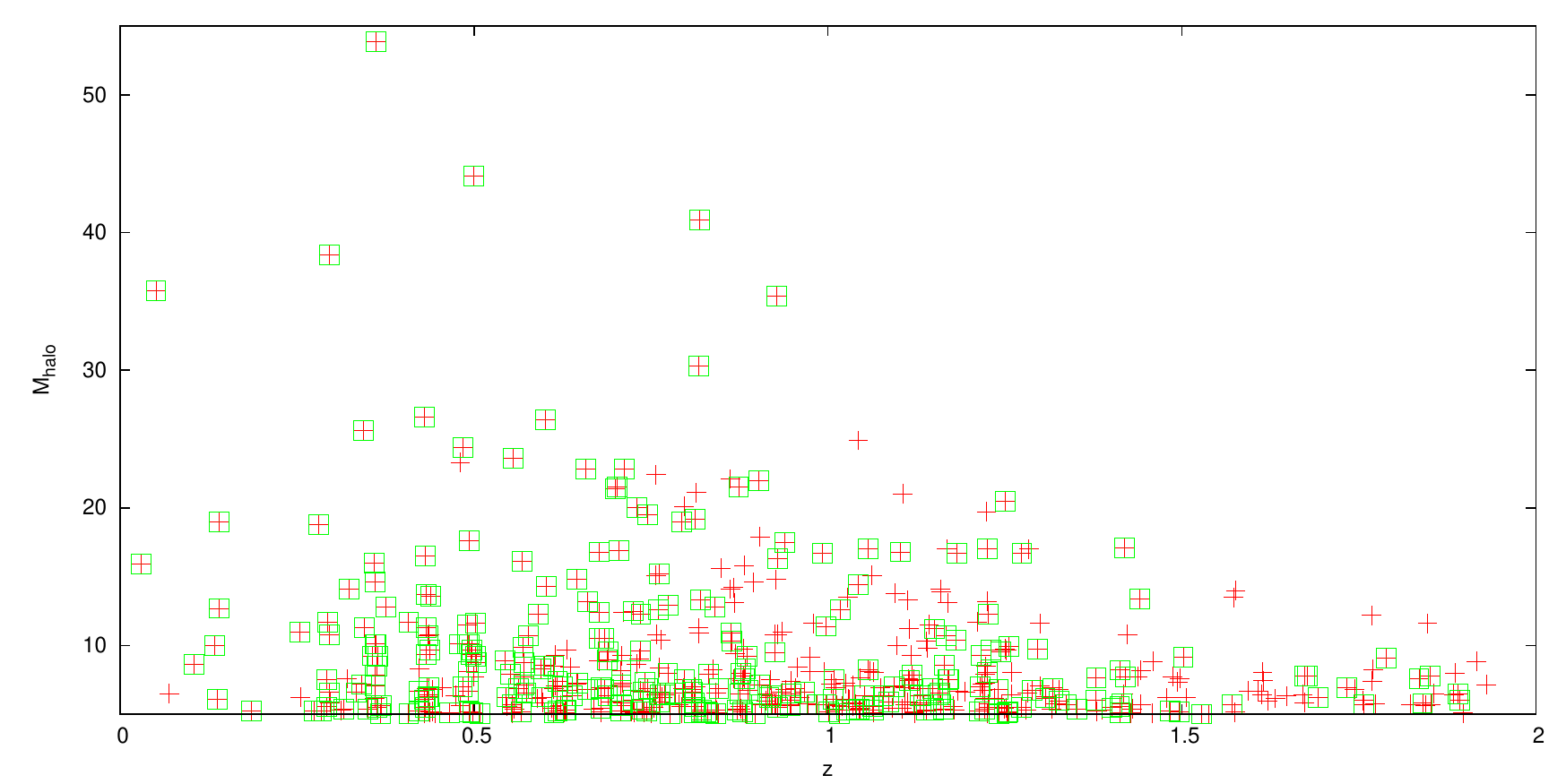}
   \includegraphics[width=7.5cm, angle=0]{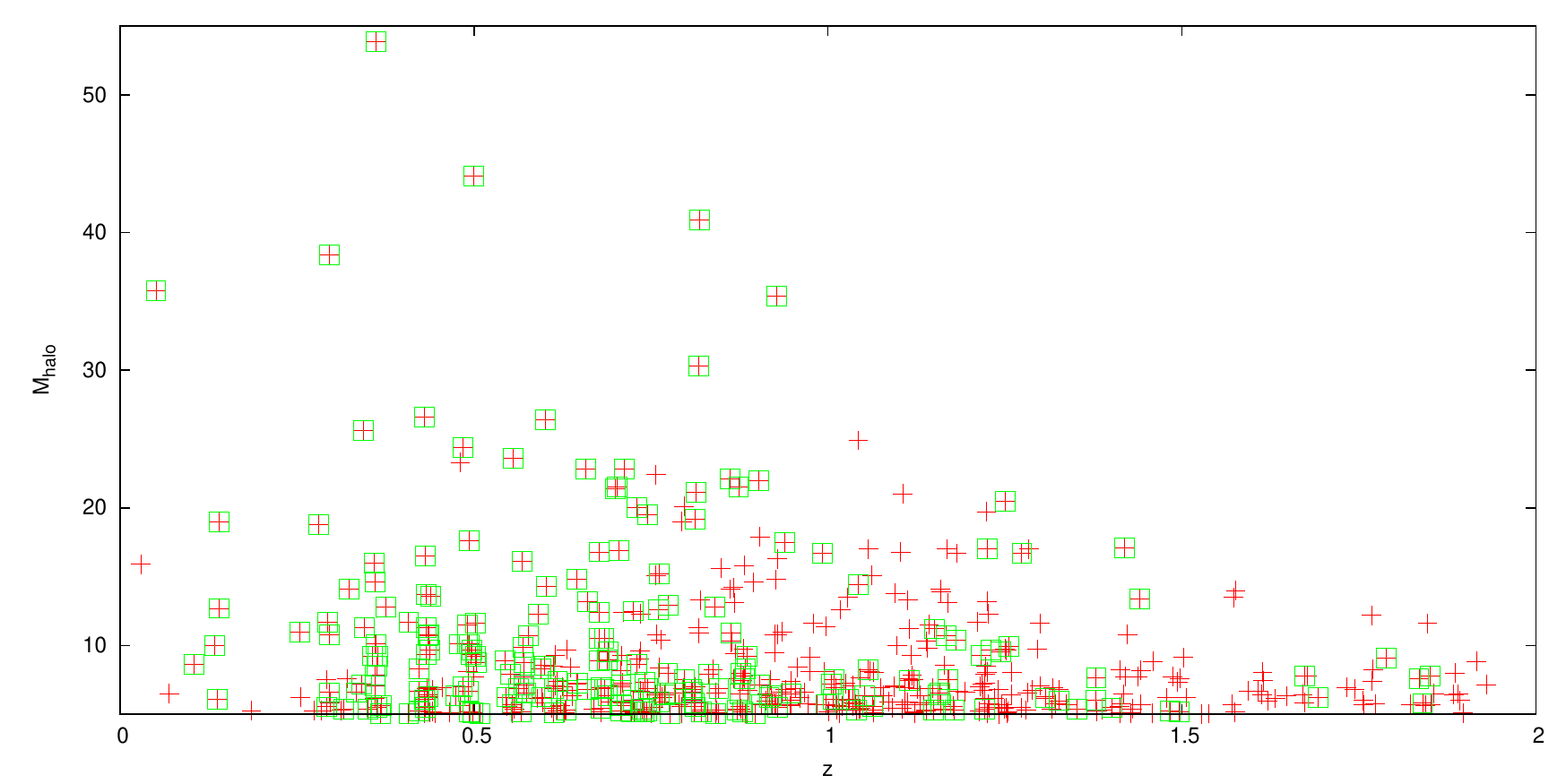}\\
   \includegraphics[width=7.5cm, angle=0]{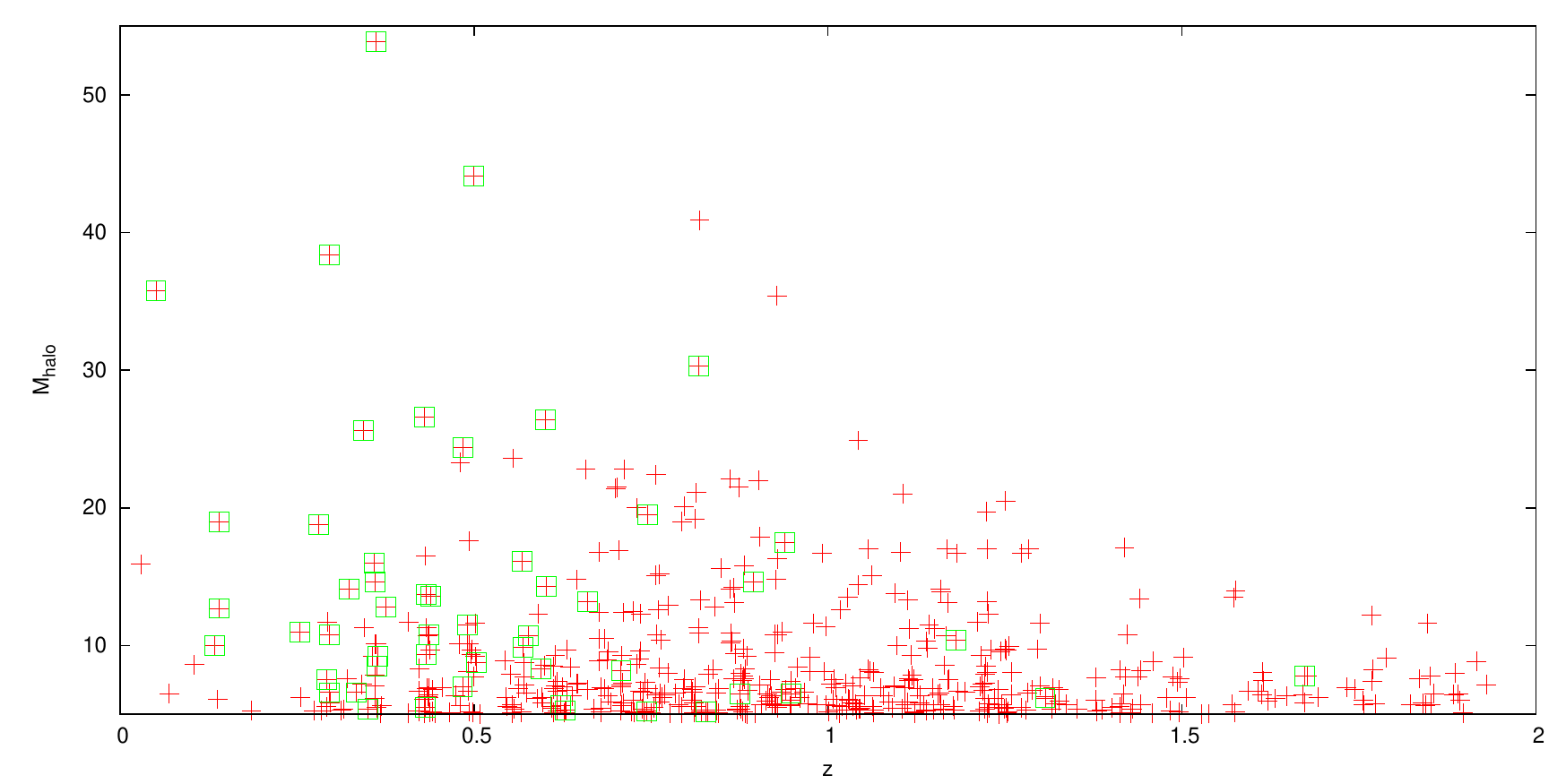}
   \includegraphics[width=7.5cm, angle=0]{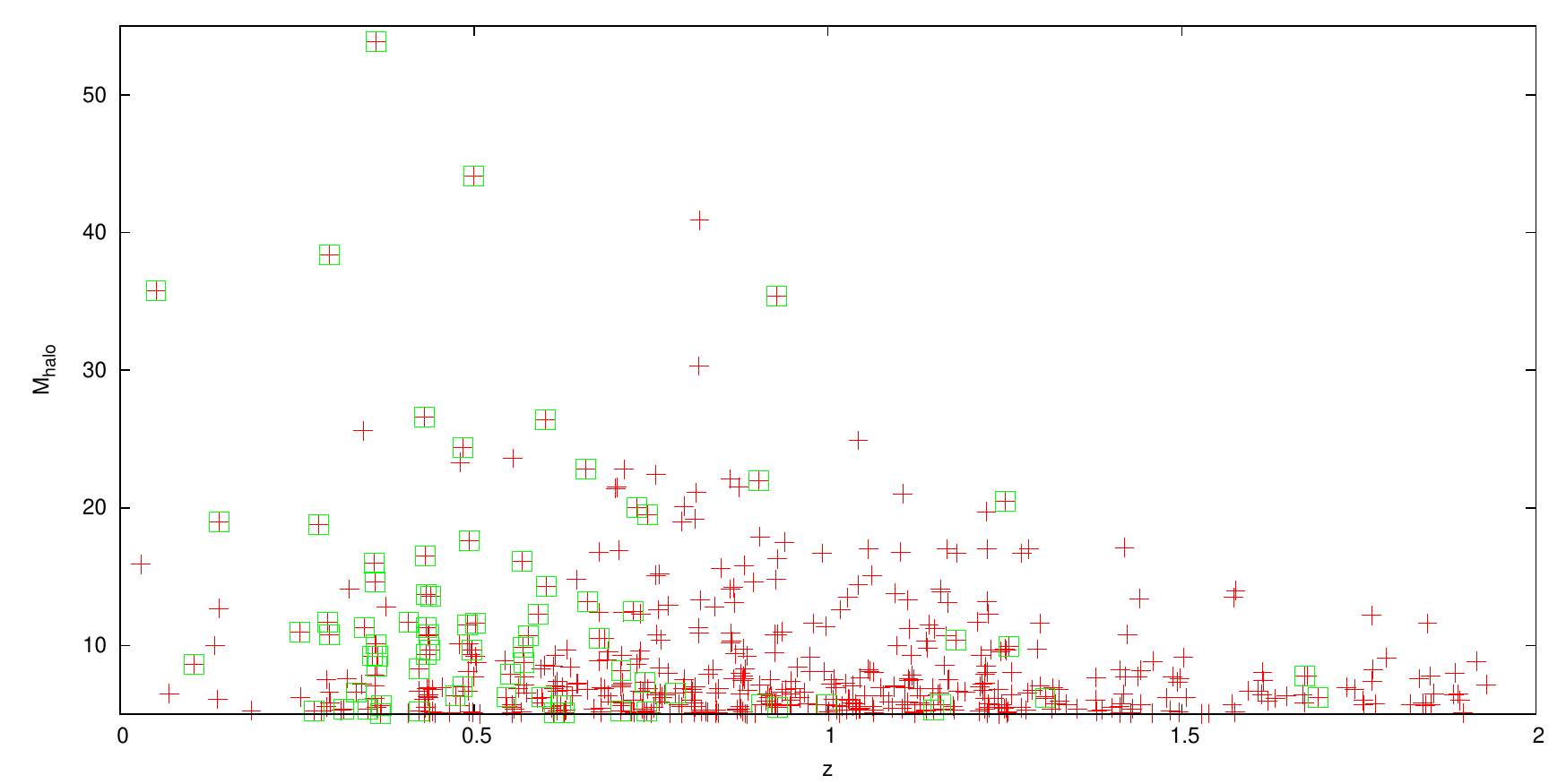}
   \caption{Peak-halo correspondences in $z-M$ plane. The upper and lower 
panels are for the Gaussian smoothing and MRLens, respectively. The left and right panels
are for $n_g=30\hbox{ arcmin}^{-2}$ and $n_g=100\hbox{ arcmin}^{-2}$, respectively.
In each panel, the `+' symbols show all the
halos with $M\ge 5\times 10^{13} M_{\odot}$ and the squares show the
halos with corresponding convergence peaks with $\kappa\ge 0.02$. }
   \label{Fig6}
   \end{figure}

In MRLens, the $\alpha_0$ parameter plays a crucial role in classifying 
significant and non-significant wavelet coefficients. A larger $\alpha_0$ leads
to a larger fraction of significant coefficients, and thus a less suppression effect 
in MRLens restoration. To see if the problem of low completeness in MRLens cluster detection 
can be largely improved by increasing $\alpha_0$, we analyze the $\alpha_0$ dependence for the completeness
$f_c$ as well as for the efficiency $f_e$. The results are shown in Figure~8.
The upper and lower panels are for $n_g=30\hbox{ arcmin}^{-2}$ and $n_g=100\hbox{ arcmin}^{-2}$, respectively.
The red, green and blue lines are for $M\ge 5\times
10^{13}M_{\odot}$, $1\times 10^{14}M_{\odot}$, and $2\times
10^{14}M_{\odot}$, respectively. The peak detection threshold is set to be $\kappa=0.02$.
We can see that both $f_c$ and $f_e$ are not very sensitive to $\alpha_0$. 
Increasing $\alpha_0$ from $0.01$ to $0.1$
improves the completeness only by $\Delta f_c\sim 10\%$ for both $n_g=30\hbox{ arcmin}^{-2}$, 
and $n_g=100\hbox{ arcmin}^{-2}$. Meanwhile, the efficiency decreases by $\Delta f_e\sim 10\% -20\%$.
Therefore increasing $\alpha_0$ cannot overcome the shortcoming of MRLens considerably. 
Comparing to the Gaussian smoothing cases, the completeness is still low for MRLens weak-lensing cluster detection
even with $\alpha_0=0.1$. 

We then conclude that in probing cosmologies with weak-lensing cluster abundance analyses, in which
a large sample of clusters is needed, the Gaussian smoothing method performs much better than the MRLens
method. To overcome the relatively low efficiency for low peaks in the Gaussian smoothing treatment,
a detection threshold $\kappa>3\sigma_0$ is normally set. In Fan et al. (2010), the noise effects
on convergence peak statistics can be accurately modeled for the Gaussian smoothing method. Therefore
it is potentially possible to even include peaks with $\kappa<3\sigma_0$ in the abundance analyses, which
can increase the number of detected clusters greatly so that to strengthen the derived cosmological constraints on different
parameters. This will be explored further in our future studies. 
 
\begin{figure}
\centering
\includegraphics[width=10.0cm]{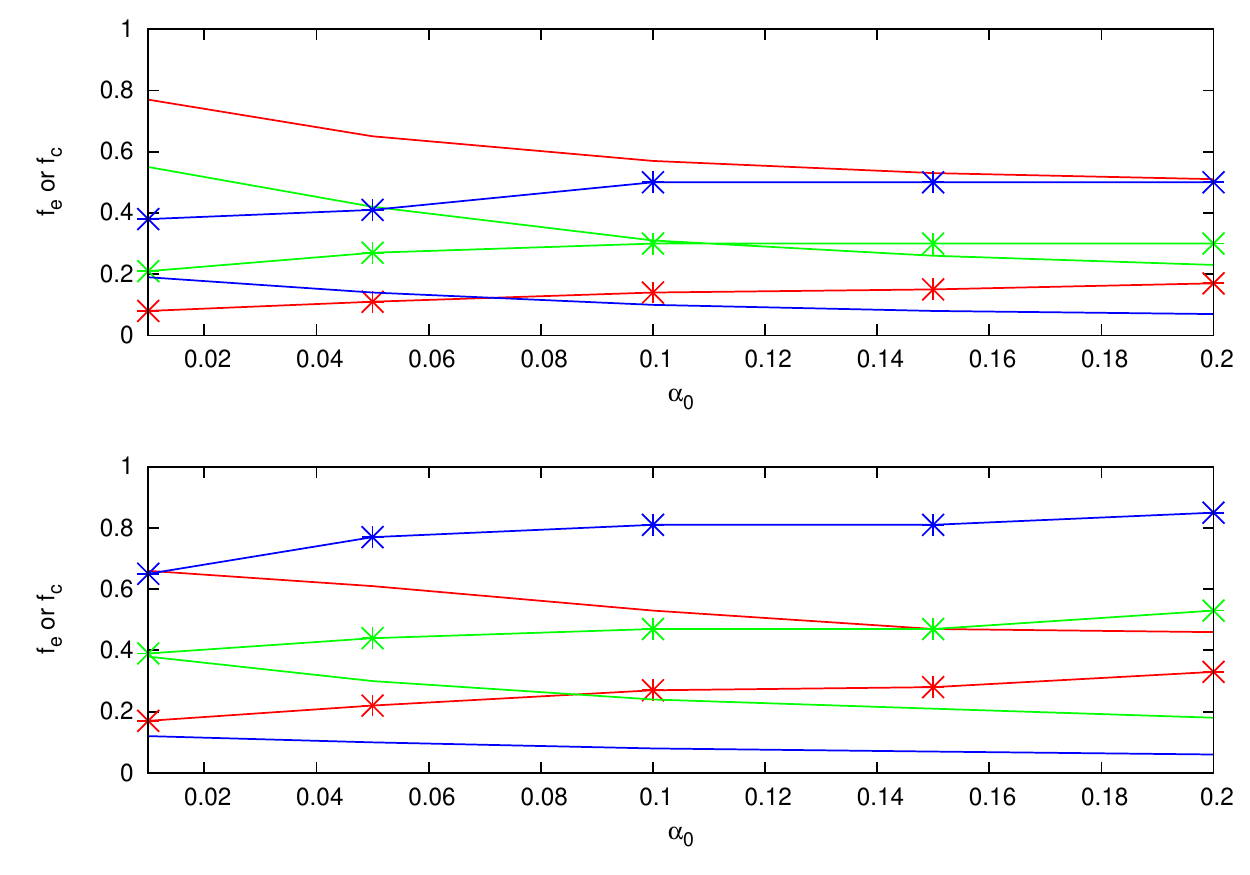}
\caption{The $\alpha_0$ dependence of the completeness $f_c$ and the efficiency $f_e$. 
The upper and lower panels are for $n_g=30\hbox{ arcmin}^{-2}$ 
and $n_g=100\hbox{ arcmin}^{-2}$ respectively. The peak detection threshold is set to be
$\kappa=0.02$. The red, green and blue lines are for the results of
halos with $M >5\times10^{13} M_{\odot}$, $M
>1\times10^{14} M_{\odot}$, and $M >2\times10^{14} M_{\odot}$,
respectively. The lines with and without symbols are for the completeness $f_c$ and the 
efficiency $f_e$, respectively.} \protect\label{fig8}
\end{figure}

\section{Summary}
\label{sect:summary}

Constructing cluster samples through their weak-lensing effects has
been an important aspect of weak-lensing studies. Their statistical
abundance contains valuable cosmological information. Observations
have shown the feasibility in detecting clusters with weak-lensing
effects (e.g., Wittman et al. 2006; Dietrich et al. 2007; Gavazzi \&
Soucail 2007; Schirmer et al. 2007; Hamana et al. 2009). In
conjunction with optical observations, the detailed analyses on the
completeness and efficiency of weak-lensing selected cluster samples
also become possible (e.g., Geller et al. 2010). It is noted,
however, the efficiency and completeness depend on the method
applied to reconstruct the convergence field from shear
measurements. Different methods can result residual noise with
different statistical properties, and can also change the
weak-lensing signals differently. In order to extract cosmological
information from observations, it is therefore crucial to understand
how a particular reconstruction method affects the results in
detail.

In this paper, we systematically compare the Gaussian smoothing method and
the MRLens treatment to suppress noise from intrinsic ellipticities
in convergence maps. We concentrate on convergence peak statistics.
It is found that while the MRLens method can remove noise very
effectively, it mistakenly removes a large fraction of real peaks associated with clusters of
galaxies. For $n_g=30\hbox{ arcmin}^{-2}$, the number of peaks with $\kappa\ge 0.02$
after MRLens filtering is only $\sim 50$ in an area of $3\times 3\hbox{ deg}^{2}$
in comparison with $\sim 530$ for the number of halos of $M>5\times 10^{13}\hbox{ M}_{\odot}$.  
On the other hand, for the Gaussian smoothing treatment, the number of detected clusters
is $\sim 260$. Even with the detection threshold $\kappa=3\sigma_0\sim 0.045$, which is normally set  
in the Gaussian smoothing treatment to reduce the number of noise peaks in the peak catalog and thus 
to increase the cluster detection efficiency, 
the number of detected clusters is $\sim 100$, twice as many as that in the MRLens filtering with the
threshold $\kappa=0.02$. As the accuracy of statistical abundance analyses depends crucially on 
the number of detected clusters, the Gaussian smoothing method is therefore strongly favored 
to detect clusters as many as possible. Furthermore, the Gaussian smoothing leads to a noise
field which is approximately Gaussian in statistics, while the residual noise from MRLens filtering
is highly non-Gaussian. Therefore the noise effects can be modeled more straightforwardly for the
Gaussian smoothing case than that of MRLens (e.g., van Waerbeke 2000; Fan 2007; Fan et al.
2010). The recent studies of Fan et al. (2010) on the weak-lensing peak statistics with noise included
provide an analytical model for the efficiency of peak detections in the Gaussian smoothing case.   
Thus it is possible for us to include peaks with $\kappa<3\sigma_0$ in the analyses.
Then the number of detected clusters can increase considerably, which in turn can lead to a 
significant improvement in the cosmological constraints derived from weak-lensing cluster statistics.

\begin{acknowledgements}
This research is supported in part by the NSFC of China under grants
10373001, 10533010 and 10773001, and the 973 program
No.2007CB815401. HuanYuan Shan is very grateful for the hospitality
of CPPM.
\end{acknowledgements}

\label{lastpage}

\end{document}